\documentclass[manuscript]{aastex}
\usepackage{lscape, morefloats, graphicx, float} 

\shorttitle{Effective Temperature Estimates for Main-Sequence Stars}
\shortauthors{Eker et al.}

\begin{document}


\title{Main-Sequence Effective Temperatures from a Revised 
Mass-Luminosity Relation Based on Accurate Properties}


\author{Z. Eker\altaffilmark{1}, F. Soydugan\altaffilmark{2,3}, 
E. Soydugan\altaffilmark{2,3}, S. Bilir\altaffilmark{4}, 
E. Yaz G\"ok\c ce\altaffilmark{4}, I. Steer\altaffilmark{5}, 
M. T\"uys\"uz\altaffilmark{2,3}, T. \c Seny\"uz\altaffilmark{2,3} 
and O. Demircan\altaffilmark{3,6}}

\affil{Akdeniz University, Faculty of Sciences, Department of 
Space Sciences and Technologies, 07058, Antalya, Turkey}

\affil{\c Canakkale Onsekiz Mart University, Faculty of Arts and 
Sciences, Department of Physics, Terzio\v{g}lu
Kamp\"{u}s\"{u}, TR-17020 \c{C}anakkale, Turkey}

\affil{\c{C}anakkale Onsekiz Mart University, Astrophysics Research 
Centre and Ulup{\i}nar Observatory, Terzio\v{g}lu Kamp\"{u}s\"{u}, 
TR-17020 \c{C}anakkale, Turkey}

\affil{Istanbul University, Faculty of Science, Department of 
Astronomy and Space Sciences, 34119, University-Istanbul, Turkey}

\affil{NASA/IPAC Extragalactic Database, Pasadena, California, USA} 

\affil{\c Canakkale Onsekiz Mart University, Faculty of Arts and 
Sciences, Department of Space Science and Technologies,\\ TR-17020 
\c Canakkale, Turkey}

\email{eker@akdeniz.edu.tr}



\begin{abstract}
The mass-luminosity ($M-L$), mass-radius ($M-R$) and mass-effective temperature ($M-T_{eff}$) diagrams for a subset of galactic nearby main-sequence stars with masses and radii accurate to $\leq 3\%$ and luminosities accurate to $\leq 30\%$ (268 stars) has led to a putative discovery. Four distinct mass domains have been identified, which we have tentatively associated with low, intermediate, high, and very high mass main-sequence stars, but which nevertheless are clearly separated by three distinct break points at 1.05, 2.4, and 7$M_{\odot}$ within the mass range studied of $0.38-32M_{\odot}$. Further, a revised mass-luminosity relation (MLR) is found based on linear fits for each of the mass domains identified. The revised, mass-domain based MLRs, which are classical ($L \propto M^{\alpha}$), are shown to be preferable to a single linear, quadratic or cubic equation representing as an alternative MLR. Stellar radius evolution within the main-sequence for stars with $M>1M_{\odot}$ is clearly evident on the $M-R$ diagram, but it is not the clear on the $M-T_{eff}$ diagram based on published temperatures. Effective temperatures can be calculated directly using the well-known Stephan-Boltzmann law by employing the accurately known values of $M$ and $R$ with the newly defined MLRs. With the calculated temperatures, stellar temperature evolution within the main-sequence for stars with $M>1M_{\odot}$ is clearly visible on the $M-T_{eff}$ diagram. Our study asserts that it is now possible to compute the effective temperature of a main-sequence star with an accuracy of $\sim 6\%$, as long as its observed radius error is adequately small ($<1\%$) and its observed mass error is reasonably small ($<6\%$).
\end{abstract}


\section{Introduction}
One of the fundamental secrets of the cosmos, the famous stellar mass-luminosity relation (MLR), was discovered empirically in the beginning of the 20th century independently by \citet{Hertzsprung23} and \citet*{Russell23} from the masses of visual binaries. Eclipsing binaries were included in MLRs later. There were only 13 eclipsing binaries available together with 29 visual binaries and 5 Cepheids to \citet{Eddington26}, while \citet{McLaughlin27} was able to include 41 eclipsing binaries in his plots.     

Investigations of the MLR continued as the quantity and quality of data increased \citep*{Kuiper38, Petrie50a, Petrie50b, Strand54, Eggen56, Cester83, Henry93}. Especially noteworthy was the critical compilation of absolute dimensions of binary components by \citet{Popper67,Popper80}. Only after the mid 20th century however, did studies involving the  empirical stellar mass-radius relation (MRR) begin to appear in the literature \citep{McCrea50, Plaut53, Huang56, Lacy77, Lacy79, Kopal78, Patterson84, Gimenez85, Harmanec88, Demircan91}. There were five resolved binaries, 14 visual binaries and 12 O-B binaries with less accuracy in the MRR by \citet{Gimenez85}. \citet{Demircan91} studied both MLR and MRR using 140 stars (70 eclipsing binaries) including the main-sequence components of detached, semi-detached binaries and the components of OB-type contact and near-contact binaries

\cite{Andersen91} collected detached double-lined eclipsing systems having masses and radii with uncertainties within 3\%. \citet{Henry93} used masses of 37 visual binaries with main-sequence components from the fourth catalog of orbits of visual binary stars \citep{Worley83} in order to study the MLR for stars of mass 0.08 to 1$M_{\odot}$ at near-infrared wavelengths, $J$ (1.25 $\mu$m), $H$ (1.6 $\mu$m), and $K$ (2.2 $\mu$m). For the mass-absolute magnitude relation at visual wavelengths, where the upper mass limit extended to $2 M_{\odot}$, \citet{Henry93} combined the visual/speckle binary sample with the eclipsing binary data of 24 systems taken from \citet{Andersen91} and \citet{Popper80}. \citet{Gorda98} collected stellar masses and radii with accuracies within 2-3\% from photometric, geometric, and absolute elements of 112 eclipsing binaries with both components on the main sequence for studying ($\log m-M_{bol}$) and ($\log m-\log R$) relations, where $m$, $R$ and $M_{bol}$ stand for stellar mass, radius and absolute bolometric magnitude. 

\citet{Ibanoglu06} studied 74 detached and 61 semi-detached Algols on $M-R$, $M-T_{eff}$, $R-T_{eff}$, and $M-L$ diagrams separately. Collecting masses, luminosities, radii and temperatures from 114 (215 stars) detached main-sequence eclipsing binaries, \citet{Malkov07} constructed MLR and other relations [$M_V(\log m)$, $\log m(M_V)$, $\log L(\log m)$, $\log m(\log L)$, $\log T_{eff}(\log m)$, $\log m(\log T_{eff})$, $\log R(\log m)$, $\log m(\log R)$] by polynomials validated for the ranges of data. Recently, \citet*{Torres10} updated the critical compilation of accurate, fundamental determinations of binary masses and radii. Studying metallicity and age contributions to the MLR for main-sequence FGK stars, \citet*{Gafeira12} worked with only 13 binary systems, all taken from \citet{Torres10}.

Mass and chemical composition are two independent basic parameters from which various stellar evolution models are constructed. Radius ($R$), luminosity ($L$), and effective temperature ($T_{eff}$) are prime products of these models, and parameters by which stellar evolution can be followed. From the observational point of view, $M$ and $R$ are first order observable parameters. In the last decade, the number of accurately determined $M$ and $R$ has increased by hundreds. Further, their accuracy has reached the level needed to indicate stellar evolution, even within the main-sequence band. Chemical composition (or metallicity $[M/H]$) and $T_{eff}$ are not as accurate as $M$ and $R$, as present data indicates for binary stars. ``Spectroscopic metal abundance determinations are available for a handful of systems'' says \citet{Andersen91}. Unlike $M$ and $R$, $L$ is a second step product obtained from either observed magnitudes and distance, or from the relation $L=4\pi R^2\sigma T^4$. Both are problematic however, because both require accurate determinations of effective temperatures and such temperatures are rarely accurate because they are indirectly inferred. This study asserts that a revised main-sequence MLR can provide an easy and effective estimate of stellar effective temperatures directly from masses and radii.

Table 1 summarizes MLRs, MRRs and related improvements within the last two decades, where contributions from detached eclipsing-double lined binaries appear dominant. According to Table 1, the number of stars first decreased from 140 (70 eclipsing pairs) to 90 by eliminating unreliable ones. \citet{Demircan91} kept contact and semi-contact systems in calculating empirical $M-M_{bol}$ (mass-absolute bolometric magnitude) and MRRs within the mass range $0.63<M/M_{\odot}<18.1$ for the sake of statistical significance. Afterwards, the number of stars increased to 188, where 149 of them were from detached eclipsing main-sequence binaries in the study of \citet{Henry04}, who also extended the mass range to $0.07<M/M_{\odot}<33$. Numbers increased further: 215 stars (114 pairs) by \citet{Malkov07} and 190 stars (94 eclipsing and $\alpha$ Cen) by \citet{Torres10}. An online database of detached binaries known as DEBCat\footnote {http://www.astro.keele.ac.uk/jkt/debcat/}\citep{Southworth14}, which is periodically updated and commonly referenced, currently includes 342 stars (171 pairs). For this study, the latest compilation of detached eclipsing binaries given by \citet{Eker14}, was chosen as a calibration sample. It features 514 stars in 257 detached double-lined eclipsing binary systems.

Table 1 makes clear that the most recent updates of MLR and MRR are from \citet{Malkov07}. It is noteworthy that despite having the most accurate and sufficient number of data, \citet{Andersen91} and \citet{Torres10} did not pursue revising the classical MLRs \& MRRs. ``The scatter on the mass-luminosity diagram is not due to observational errors but most likely abundance and evolutionary effects'' was a genuine message in both studies. \citet{Andersen91} claimed ``...  departures from a unique relation are real''. Therefore,  declining to define a unique $M$-$L$ function, \citet{Torres10} preferred to express $\log M$ by the simplest possible polynomials containing the parameters $T_{eff}$, $\log g$ and $[Fe/H]$ as variables rather than a classical approach as \citet{Malkov07}, who expressed $\log L$ as a function of $\log M$. Similarly, a unique MRR was not defined by \citet{Torres10} but instead $\log R$ was expressed as a function of $T_{eff}$, $\log g$ and $[Fe/H]$, unlike in  \citet{Malkov07}, who expressed $\log R$ ($\log M$) and/or vice versa $\log M$ ($\log R$). 

Fortunately, it is always possible to define a unique function to express a band of data by a least squares method. Moreover, the classical MLR and MRR have been proven to be useful and are widely applied. For example, the MLR and MRR of \cite{Malkov07} has been used by \cite{Catanzaro14}, \cite{Fumel12}, \cite{Ripepi11}, and \cite{Zasche11}. The MLR and MRR of \cite{Demircan91} were used by \cite{Jiang09}, \cite{Kraus09}, \cite{Hunter08}, and \cite{Li05}. Of course, usage is not limited to these authors. There are many more examples; indeed too many for all to be cited here. Therefore, revisions of classical MLR and MRR with modern and more accurate data are periodically necessary. As a preparatory work, \citet{Eker14} have already compiled basic stellar parameters (i.e. $M$, $R$, $T_{eff}$ and $L$) for 257 detached binaries (514 stars), mostly main-sequence, and with masses and radii recomputed based on the most up to date values in use of the solar gravitational constant and solar radius of $GM_{\odot}=1.3271244\times10^{20}$ m$^{3}$s$^{-2}$ \citep{Standish95} and $R_{\odot}=6.9566\times$10$^{8}$m \citep*{Haberreiter08}. Stellar temperatures and luminosities were also revised and homogenized \citep[Table 2 of][]{Eker14}. Compared with previous lists in Table 1, the number of stars with masses and radii, which are accurate to within 3\%, is increased by 50\%. Not only the number and quality, but also the range of the data was improved to: $0.2<M/M_{\odot}<32$, and $0.23<R/R_{\odot}<9.36$. All of these improvements motivated us to revise the $M-L$, $M-R$ and $M-T_{eff}$ relations based on the increased quantity and quality of data, and the fact that it is now available in a homogenized form.  

Critically, most light curve solutions for eclipsing binaries require an effective temperature for at least one component as an input. However, in most critical cases that is not available. Therefore, some authors such as \citet{Helminiak09} and \citet{Zasche11b}, avoid using roughly estimated values, and are satisfied with solutions without temperatures. Such solutions contain only temperature ratios, not absolute temperatures. Since detached double-lined eclipsing binaries are basic sources of reliable stellar parameters ($M$, $R$), this study asserts that by revising the MLR, and making use of $L=4\pi R^2\sigma T^4$, direct estimates of stellar temperatures can be obtained. Direct estimates can be used in problematic cases where no estimate is available, and can provide an important, independent calibration of temperatures obtained by indirect means.

\section{The Data}
The number of stars with dependable parameters was so limited just a quarter of a century ago, that even solutions based on contact and semi-contact eclipsing systems could not constrain uncertain parameters sufficiently to produce revised MLRs and MRRs, as shown in Table 1. Thanks to modern detectors, observing techniques and high speed computers, the number of accurate and precise solutions from detached eclipsing spectroscopic binaries has increased rapidly. Today we are able to select stars with the most reliable stellar parameters of any desired criteria. Therefore, the first step of this study was to form a calibration sample containing a larger number of stars with more reliable parameters distributed over a greater range than was available to previous studies (Table 1). 

\subsection{Calibration Sample}
A calibration sample was formed by selecting main-sequence stars with the most accurate masses, radii and effective temperatures from Table 2 of ``The Catalogue of Stellar Parameters ...'' by \citet{Eker14}, which is already reprocessed and homogenized. In the first step, our preliminary criteria involved finding stars where both mass and radius with errors of less than or equal to 3\%, and luminosities with errors less than or equal to 30\% were available. Among 514 stars (257 binaries), 296 stars were found fulfilling the criteria. In the second step, 25 stars outside of the main sequence were removed.
  
The process of removing non-main-sequence stars was completed by using the mass-radius diagram. Compared to effective temperatures and luminosities, which can only be inferred indirectly, masses and radii provide much more reliable indicators of stellar properties, and a highly improved diagnostic tool for analyzing stellar evolution. Fig. 1 shows 271 main-sequence stars selected for the calibration sample and 25 non main-sequence stars on the $M-R$ diagram. Theoretical ZAMS (Zero Age Main Sequence) and TAMS (Terminal Age Main Sequence) lines for metallicity zero from \citet{Bertelli08, Bertelli09} were used as border lines to secure the stars within the main-sequence band.
  
Although metallicity data is missing in the catalogue of \citet{Eker14}, the thin-disk field stars in the solar neighborhood are known to have solar metallicity on average, with upper and lower limits of $\pm0.5$ dex in metallicity  \citep{Cox00}. It is also known that the thin disk stars in the solar neighborhood could be polluted by about 6-8\% by thick disk and halo stars \citep{Karaali03, Karatas04, Bilir05, Bilir06, Bilir08, Ak10, Ak13}. TAMS lines for zero metallicity, therefore, do not limit the upper border sharply. Consequently, there could be a negligible number of non-main sequence stars remaining in the calibration sample. Such a small amount of pollution however, is too small to alter the general characteristics of the sample stars, which are solar neighborhood main-sequence stars with an average metallicity of zero. 

Table 2 gives basic astrophysical parameters ($M$, $R$, $T_{eff}$, and $L$) and their relative errors. The columns are self explanatory to indicate sequence number, name, coordinates, component (primary or secondary), mass, relative error in mass, radius, relative error in radius, published effective temperature, error in temperature, luminosity, relative error in luminosity, the Roche lobe filling factor and remark. A filling factor (FF) for a star in a binary, which is defined as $FF={\bar{r}}/{\bar{r}}_{RL}$ , where ${\bar{r}}$ is average radius, and ${\bar{r}}_{RL}$ is Roche Lobe radius relative to the semi-major axis of the orbit, is a parameter that indicates its sphericity. \cite{Eker14} have concluded that deviations from sphericity could be ignorable for small $FF$, as big as 75\%, a value corresponding to the difference between $r$(point) and $r$(pole) being less than 1\% of the star radius. $r$(point) and $r$(pole) are the deformed radii of the star towards the other component and towards the rotation axis, respectively. 

Fig. 2 shows how filling factors distribute among the stars of the calibrating sample. Accordingly, 89\% of the stars in the calibrating sample are spherical within 1\% of radius. The rest, or 11\%, are deformed more than 1\% but still respectively detached, so that one can assume all of the calibrating stars are from binaries in which mass transfer has not occurred. All of the stars in the calibrating sample should have evolved as if they were single with a corresponding rotation as indicated in the catalogue of \cite{Eker14}.

\subsection{Stellar mass domains}
The main sequence MLR, discovered by \citet{Hertzsprung23} and \citet{Russell23} in the first half of the 20th century, is one of the most fundamentally confirmed, and universally recognized astronomical relations. The relation can be expressed in different ways, based on the relation between $M$ and $L$, or between $M$ and $M_{bol}$ for example. Various forms of it are found.  However, the most fundamental and basic form is $L \propto M^{\alpha}$, where the power ($\alpha$) is the slope on the logarithmic $M-L$ diagram. With early limited data, it was possible to express the MLR of main-sequence stars with a single power law. However, as the quantity, quality, and range of data increased, it became increasingly clear that the MLR could not be expressed with a single linear fit. Consequently, many authors \citep{Kuiper38, Cester83, Andersen91, Demircan91, Henry93, Malkov03, Malkov07, Fang10} preferred to express the MLR for various mass ranges; thus there could be various values of $\alpha$ for various mass ranges, usually arbitrary.

When pre-exploring different forms of the relation using $M$ and $L$ of 271 stars (Table 2), we identified natural stellar mass intervals, where $\alpha$ is constant on the $\log M-\log L$ diagram. Their identification came as a surprise, since we were exploring the amount of stellar energy production rate as a function of stellar mass. Being a parameter expressing the efficiency of stellar furnaces, that is $L/M$, the luminosity per stellar mass (in solar units) was plotted against the stellar mass in Fig. 3. The distribution appears linear only in distinct mass ranges. There are break points as indicated on the figure at $M=1.05$, 2.4 and 7$M_{\odot}$. Linear distributions are clearly visible before, between and after the break points. 

The reciprocal of $L/M$, that is $M/L$, which is known as the mass to light ratio, is a parameter commonly used among extragalactic astronomers \citep{Faber10, Bell01, Girardi02}. In global meaning, mass to light ratio implies a variety of galactic compositions (stellar) and depends on the relative number of stars of different types. As it is used for spiral galaxy rotation curve decomposition and band-pass dependent slope of the galaxy magnitude-rotation velocity relation \citep[Tully-Fisher relation,][]{Tully77}, it is also recognized as a parameter to indicate dark matter \citep{Blumenthal84}. Although usage of $M/L$ by extragalactic astronomers appears to be conceptually different than $L/M$ investigated in this study, $M/L$ cannot be independent of $L/M$. We believe updated information of the energy generation efficiency as a function of varying stellar types on the main sequence will be used in improving models for future extragalactic studies.     

We tried many simple functions to express $L/M$ in terms of $M$. However, when we plotted $\log (L/M)$ versus $M$, the appearance of natural stellar mass intervals caught our attention. In Fig. 3 for example, a sharp linear increase of $\log (L/M)$ up to $1.05M_{\odot}$ is clear. For stars with masses greater than $1.05M_{\odot}$ however, that increase continues less steeply up to $2.4M_{\odot}$. The p-p chain is the main energy source for stars less massive than the Sun, while the CNO cycle becomes dominant for stars more massive than the Sun. Thus, we surmise that the break point at $1.05M_{\odot}$ is just an indication of this change. There could be similar reasoning related to the efficiency of stellar energy production mechanisms at the other break points, where the rate of linear increase of $\log(L/M)$ suddenly decreases. The preliminary conclusion, based on Fig. 3, is that the change of energy production rate per stellar mass (efficiency, $ L/M$) for stars of a given mass is a stronger function of a star's mass than of it's temporal evolution. Nevertheless, the prime concern of the present work is to study classical MLR in general and/or in between those break points, which can be defined in terms of stellar mass domains: low mass ($0.2<M/M_{\odot}\leq 1.05$), intermediate mass ($1.05<M/M_{\odot}\leq 2.4$), high mass ($2.4<M/M_{\odot}\leq 7$), and very high mass ($M/M_{\odot} >7$). We encourage theoreticians especially nuclear astrophysicists, to further investigate the physical facts and reasoning behind these break points. 

\section{Calibrations}
\subsection{Classical MRL}
The improved data of the present study permits us to apply a classical approach thanks to the recognition of stellar mass domains when updating classical MLRs. Recent studies, by comparison, were more limited \citep{Demircan91, Henry93, Malkov03, Malkov07, Fang10}. At best, the existing mass range could be divided at $M=1.7M_{\odot}$, since stars of $M>1.7M_{\odot}$ have convective cores with radiative envelope while stars of $M<1.7M_{\odot}$ are vice versa. At worst, the range in masses was handled arbitrarily when determining the power of $M$ for the classical MLRs. Fig. 4 gives the distribution of the sample stars on the $\log L-\log M$ diagram. The break points in Fig. 3 are marked as vertical dashes at 0.021, 0.38 and 0.845 in $\log M$, which corresponds to 1.05, 2.4 and $7M_{\odot}$ in mass. The first one at $1.05M_{\odot}$ is obvious but other break points are not as clear as before when compared to Fig. 3, where the mass axis was linear. With a careful look at Fig. 4, one can nevertheless still detect by eye the linear orientation of the data before, between and after these break points, even with the mass axis changed to logarithmic and vertical axis changed to $\log L$. 

The linear distributions of $\log L$ within the four mass domains are clearly displayed in the four panels of Fig. 4 (Fig. 4b, c, d and e) below the main panel (Fig. 4a). The four panels contain $M$ and $L$ of the stars from the calibrating sample within each stellar mass domain, which was defined according to the brake points in Fig. 3. The classical MLR ($L \propto M^{\alpha}$) for stars within each of the four stellar mass domains have been determined by fitting a linear equation by the least squares method. The statistics, mass domains and  linear MRLs are summarized in Table 3. The linear equations given in column 4 represent the best fitting lines, which are displayed in the lower four panels in Fig. 4. High degrees of correlations indicated by the data within the lines of $1\sigma$ limit are very clear. Notice that the three stars with the smallest masses, the components (A and B) of CM Dra and the secondary of LSPM J1112+7626, do not seem to obey the linear trend of the MLR line in the low mass domain ($0.2<M/M_{\odot}\leq1.05$). Therefore, those three stars were excluded from the analysis and the lower limit of the low mass domain was changed to $0.38M_{\odot}$ as indicated in Table 3. Obviously, those three stars with lowest masses in the calibration sample belong to another domain, which could be called very low mass domain. The break point between the very low and low mass domains is not clear because of insufficient data. We leave clarification of this point for the future, when more data will be available.       

As in earlier studies, we also asked if there is a single unique function to represent all stars in the calibrating sample. Consequently, a linear, a quadratic and a cubic equation were fitted to the masses and luminosities of the calibration sample using the least squares method. The results are summarized in Table 4. The correlations of all three fits are the same, but the linear function has considerably larger standard deviation than the quadratic and cubic functions. It is interesting that quadratic and cubic functions give the same standard deviations. This means that a quadratic function, as given in Table 4, fits the present sample sufficiently and there is no need for cubic and higher order polynomials. The quadratic MLR found in this study is compared with the most recent quadratic MLRs in Fig. 5. While deviation from \citet{Demircan91} is apparent, there is very little difference between \citet{Malkov07} and the present study.

As in earlier studies, it could be practical to define a single function to represent the MLR for stars in the calibration sample as a whole. However, if such a function is a quadratic, cubic or a higher order polynomial, the  classical meaning ($L \propto M^{\alpha}$) implied by a linear fit is lost. In revising the classical MLR based on the calibration sample, the linear fits found within the four mass domains can be considered as due to physically real if not yet physically understood divisions. Quadratic MLRs by comparison are only useful for inter-comparing the results of current and previous studies in terms of how improved the quantity and quality of data is compared to previous studies. In this study we assert therefore, that the linear MLRs as suggested in Table 3, are best to represent the calibration sample in particular and, nearby main-sequence field stars in general.    

The residuals due to the four piece linear MRL (Table 3) and residuals due to quadratic and cubic MRL (Table 4) are inter-compared in Fig. 6. The quality of all fits is very similar, but the four-piece linear function, with physical background, is preferable.

\subsection{Mass-Radius and Mass-Effective Temperature}

The MRR and the mass-effective temperature (MTR) relations found from the present sample are displayed in Fig. 7. The radius evolution within the main-sequence band, especially for the stars $M>1 M_{\odot}$, is clearly visible on the $M-L$ plot. The appearance of data on the $M-R$ diagram is very different than the appearance on the $M-L$ diagram, which rather looks like a band of data expressible by a function. However, with a very narrow distribution of radii for masses $M<1 M_{\odot}$ and a broad band of radii for stars with $M>1 M_{\odot}$, a single function to express a MRR would be odd and meaningless. A continuous line in Fig. 4a indicates the theoretical ZAMS according to  \citet{Bertelli08, Bertelli09}. However, the temperature evolution within the main-sequence band is not that obvious on the $M-T_{eff}$ diagram. At first look, it resembles the MLR shown in Fig. 4a. Therefore, one may think, it is possible to express a MTR by a polynomial or by various linear fits as was done for the MLRs of the four mass domains.

The temperatures of stars are determined mostly by a few methods including intrinsic colors, atmosphere modeling, spectral fitting to selected spectral line(s) or region(s), and/or from spectral line depth ratios. When applied to binary stars, however, those methods face severe difficulties since colors and spectra obtained are usually for the system, not for the components separately. Recent methods, like CCF fitting or spectral analysis give accurate, and reliable temperatures, but it is only quite recently. In many previous studies, including light curve analyses of eclipsing binaries, the temperatures of the primary components were adopted according to roughly estimated spectral types and colors determined from de-reddened $UBVRI$ photometry \citep*[e.g.][]{Ren11,  Li13, Elkhateeb14}. Spectral types and colors for eclipsing binaries however, are mostly given for the system, not for the components separately. In addition, low resolution spectra do not provide reliable spectral types. As a result, some researchers have preferred to use temperature ratios to avoid unreliable temperature values \citep*[e.g.][]{Helminiak09, Zasche11b}. Therefore, practical and  reliable methods for determining component temperatures in eclipsing binaries are needed.

\subsection {Calculating $T_{eff}$ using MLRs}

An effective temperature for a star may be calculated using the well known Stephan-Boltzmann relation: $L=4\pi R^2\sigma T_{eff}^4$. Solving it for $T_{eff}$
\begin{eqnarray}
T_{eff}=5777\times \sqrt[4]{\frac{L/L_{\odot}}{(R/R_{\odot})^2}},
\end{eqnarray}
where 5777 K is the effective temperature of the Sun \citep{Cox00}. Only the $L$ and $R$ of the star are needed. The solar luminosity and radius ($L_{\odot}$ and $R_{\odot}$) are also needed if $L$ and $R$ are not in solar units. Assuming $M$ and $R$ of a star are available together with relative errors, one may use the proper MLR in Table 3 to compute $L/L_{\odot}$ for the full range of stellar masses of the calibration sample of this study $0.38<M/M_{\odot}<32$. After calculating the effective temperature of a star using Eq. (1), the accuracy would be estimated as following:
\begin{eqnarray}
\frac{\Delta T_{eff}}{T_{eff}}=\sqrt[]{\Biggl(\frac{\Delta L}{4\times L}\Biggr)^2+\Biggl(\frac{\Delta R}{2\times R}\Biggr)^2}.
\end{eqnarray}
The vector form of Eq. (2) is $\frac{\Delta L}{L}=2\frac{\Delta R}{R}+4\frac{\Delta T_{eff}}{T_{eff}}$, which may be obtained from $L=4\pi R^2\sigma T_{eff}^4$ by a proper differentiation. The relative uncertainty of the radius was assumed to come from observational random errors. On the other hand, the relative uncertainty of the luminosity comes from dispersions of $L$ on $M-L$ diagram. Using the logarithmic differentiation rule $\Delta \log L=(\log e)\frac{\Delta L}{L}$ and making the standard deviations ($\sigma$) equal to $\Delta \log L$, then the relative uncertainty of the luminosity is
\begin{eqnarray}
\frac{\Delta L}{L}=\frac{\sigma}{0.4343}.
\end{eqnarray}

For a star of a given mass, the standard deviations and corresponding relative uncertainties are summarized in Table 5. The columns are self explanatory to indicate stellar mass domains, mass ranges, standard deviations, and corresponding uncertainties of the luminosities ($\Delta L/L$) in the first four columns. One fourth of $\Delta L/L$ and half of $\Delta R/R$ are in columns five and six. Since the largest uncertainty of a stellar radius is 3\% in the calibration sample, the $\Delta T/T$ in column seven is an upper limit. Actual $\Delta T/T$ could be as small as one fourth of $\Delta L/L$ depending on the value of the relative errors associated with the radius of the star.

According to Table 5, relative error of a radius ($\Delta R/R$) contributes little to the uncertainty of the effective temperature ($\Delta T/T$). The uncertainty of the predicted luminosity ($\Delta L/L$) dominates. The standard deviations from  MLRs are not the results of individual relative errors associated with observed parameters of the stars. In fact, we have chosen  stars in the calibration sample with relative luminosity errors of 30\% or less. Assuming a uniform error distribution, this would have given us a mean uncertainty on the MLRs of less than 15\%. However, the predicted relative uncertainties in Table 5 (column 4) from the standard deviations are nearly twice as large. This would clearly indicate that the standard deviations from MLRs are affected more from the natural dispersions within a band defined by ZAMS and TAMS luminosities. `` ... scatter is highly significant and not due to observational uncertainties'' say \cite{Torres10}. Metallicity is also contributing by shifting ZAMS and TAMS. Negligibility of observed errors in comparison to metallicity and evolutionary effects on $M-L$ diagram are already confirmed by \cite{Andersen91}, \cite{Torres10} and even on the observed H-R diagram by \citet{Eker14}. Indeed many of the error bars of individual points in Fig. 4 are much smaller than the printed symbols.

Because uncertainties of MLR luminosities dominate over errors associated with observed radii according to the data in Table 5, one may take it as an advantage to tolerate less accurate stellar radii. Increasing the relative error of radius from 3\% to 6\%, the uncertainty of the predicted temperature would still be less than 8\%, except for the high mass and very high mass domains, which would be extended to 10\%. 

According to Table 5, the method of computing $T_{eff}$ using present MLRs would tolerate less accurate radii, but what about the tolerance associated with the mass of the star? The following equation could be used to propagate the uncertainty associated with the MLR luminosity to the mass of the star as
\begin{eqnarray}
\frac{\Delta L}{L}=\alpha \frac{\Delta M}{M}
\end{eqnarray}
where the values of ($\Delta L/L$) and $\alpha$ are given in Table 5. Note that $\alpha$ is available only for MLRs with classical mass luminosity relation ($L\propto M^{\alpha}$). For the stellar mass domains defined in this study, the propagated uncertainty to the mass of the star is computed and recorded in the last column of Table 5. For stellar masses up to $M=2.4M_{\odot}$, about 6\% error in mass is tolerable. Tolerance level increases to 8\% and then to 13\% as the mass of the star increases to 7 and then up to $32M_{\odot}$, respectively. Consequently, it can be concluded that: when predicting the luminosity for a star of a given mass from the revised classical MLR in this study, the observational errors associated with  mass could be tolerable up to 6\% for the stellar masses up to $2.4M_{\odot}$ and even up to 10\% for stars with larger masses as indicated in the last column of Table 5. 

The method is still applicable to stars with less accurate mass and radius. With a mass having less accuracy, one must propagate the uncertainty of the mass back to the luminosity using Eq. (4). Such a propagated uncertainty ($\Delta L/L$) is expected to be poorer than the uncertainty ($\Delta L/L$) estimated from the standard deviations on MLRs through Eq. (3). Otherwise one must compare those propagated and MLR uncertainties of $\Delta L/L$ and then must insert the worst one in Eq. (2). The MLRs and the standard deviations predicted in this study from the most accurate stellar parameters ($\Delta L/L \leq30$\%, $\Delta M/M \leq3$\%, $\Delta R/R\leq 3$\%) reveal that it is possible to calculate the effective temperature of a star with an accuracy better than 8\% if the mass and the radius of the star have accuracies up to 6\%. With a negligible radius error, the temperature error could be reduced to as low as 6\%. For sun-like stars, that means 300-400 K. Such accuracy is lower than, admittedly, the accuracy available using other methods including spectroscopy, line depth ratios, template fitting, cross correlation or non-LTE spectral synthesis, at typically 100-200 K. However, an effective temperature with few (or several) hundred degrees uncertainty is very useful in cases where such information is not given, like \cite{Helminiak09} and \cite{Zasche11b}. Further, the method based on the Stephan-Boltzmann law, identified mass domains and revised MLR has the advantage over other methods of being practical and easy to apply if $M$ and $R$ are available. In comparison, other methods suffer from problems with de-reddening, issues with decomposition, and various other complexities common with spectroscopic techniques.

\section{Applications and results}
\subsection{Comparison with published temperatures and errors}
Our simple method for calculating effective temperatures based on the Stephan-Boltzmann law has been tested using updated MLRs. Temperatures calculated were compared with temperatures published. For the low mass domain, the revised MLR is based on 57 stars in the calibration sample, since we excluded the three lowest mass stars already mentioned as outliers from our analysis (primary and secondary of CM Dra and the secondary of LSPM J1112+7626). The low mass MLR therefore applies to main sequence stars in the mass range $0.38<M/M_{\odot}\leq 1.05$. Fig. 8 shows calculated (vertical) and published (horizontal) temperatures for 268 stars in the calibrating sample. The mean standard differences ($\sqrt{\langle(T_{cal}-T_{pub})^2\rangle}$) for the temperature ranges 2750-5000 K, 5000-10000K, 10000-15000K and 15000-43000K are marked as dashed lines in Fig. 8b. Those standard differences were compared to the standard deviations of calculated (mean calculated error) and published (mean published error) temperatures in Table 6.

The stellar Mass-Luminosity relation is a well determined universal law discovered in early 20th century and since later confirmed by stellar structure and evolution theory. The empirical $M-L$ diagrams produced within the last two decades indicate observational errors have little contribution to the true shape of the luminosity distribution (thickness of the band) for main-sequence stars. The width of MLRs is mostly affected by metallicity and evolution \citep{Andersen91, Torres10}. In this respect, our computed effective temperatures must be independent from the published temperatures because: 1) the uncertainty contribution of the dispersion on M-L diagrams is a dominant factor and it is mostly due to evolution and metallicity, not because of observational errors. Consequently, 2) observed temperature errors do not propagate back to calculated temperatures because they have negligible contribution to the dispersion.

The results in Table 6 indicate that mean calculated errors are about the same order as the mean calculated differences. This further supports the temperature calculating method used here. It also demonstrates that the calculated temperatures are all in close agreement with the published temperatures obtained by other methods, most based on optical photometry (color or brightness temperatures) and some involving spectroscopic techniques (excitation, ionization, or kinetic temperatures).

The results in Table 6 also indicate that mean published errors are about three times smaller than mean calculated errors. That is, a significant fraction of the published temperatures are underestimated. Underestimated temperature errors are very obvious in some studies, e.g. the effective temperatures of AE For $T_{eff}(sec)=4055\pm6$ K \citep{Rozyczka13}, XY UMa $T_{eff}(sec)=4125\pm$7K \citep{Pribulla01}, and DV Psc $T_{eff}(sec)=3614\pm8$ K \citep{Zhang07}. These stars are not the only examples. Most light curve solutions require a temperature for a component, and then solutions provide a temperature and its uncertainty for the other component. The internal temperature errors given by such light curve solutions, usually, are not realistic, but underestimated.  Occasionally, the same internal errors are assumed for both temperatures \citep*{Ribas99, Clausen10, Kraus11}. There are 31 systems, which temperature errors of the secondary is identical to the temperature error of the primary, among the 45 detached binaries listed by \cite{Andersen91}. Similarly, among the 95 binaries of \cite{Torres10}, the number of such systems is 67. 

The present sample is very heterogeneous in the sense that Table 1 of \cite{Eker14}, from where they were taken, lack a common treatment. Temperatures given in older papers are based on only rough fits.  Those given in many recent studies, by comparison, are much more rigorously derived. To compare the agreement between the calculated and published effective temperatures in terms of older versus recent papers, we have plotted published temperatures from the last seven years with a different symbol and color. Note that the number of recently determined temperatures dominates. 46\% of temperatures in the calibration sample (124 stars) were taken from papers published in the last seven years. The better agreement between the calculated temperatures and those published recently compared to those published less recently is shown in Fig. 8.

For a better comparison between the calculated and published temperatures, the photometric distances were computed for the limited number (20) of  binaries with most reliable $Hipparcos$ parallax within 100 parsecs to avoid interstellar reddening. The method of computing can be summarized as: first luminosity of each component was calculated using its radius and effective temperature. The luminosities were transformed to bolometric absolute magnitudes. Bolometric absolute magnitudes were, then, transformed to visual absolute magnitudes with proper bolometric corrections \citep{Cox00} corresponding to the temperatures used. The visual absolute magnitudes of the two stars (components) were combined to find absolute visual magnitude of the binary itself. From the apparent and absolute visual magnitudes (distance modulus), the photometric distances were computed and compared to the $Hipparcos$ distances of the selected binaries in Fig. 9. The standard deviations of the differences from the diagonal indicate that calculated temperatures are slightly less accurate than the published temperatures.

\subsection{Accuracy and Utility} 
Figure 10 compares both $M-T_{eff}$ diagrams with published and calculated temperatures. Note that the $M-T_{eff}$ diagram with published temperatures is shifted up in vertical scale by +0.3 dex in order to compare both on a single diagram. Heterogeneity of the published temperatures and homogeneity of the calculated temperatures can be deduced. The published temperatures appear mostly contained within the main-sequence band for stars with $M>1M_{\odot}$, while for the same stars the calculated temperatures appear to be scattered more and outside the main-sequence band. By comparison, the homogeneity and larger uncertainties makes the distribution of calculated temperatures relatively thicker than published ones (see, Fig. 10).

The classical MLR is a thin, well defined function, while the main-sequence evolution is rather a band. Therefore, calculating effective temperatures using such a function propagates the half thickness of the band as a dominant uncertainty, more than at least 6\%. Thus there are more points outside the ZAMS and TAMS lines of \cite{Bertelli08, Bertelli09} for the stars with $M>1 M_{\odot}$. In comparison, for stars with $M\ll 1M_{\odot}$, the main-sequence lifetimes are much greater than the age of our Galaxy. Evolutionary effects therefore, do not impact lower mass stars as greatly as they do higher mass stars. MLR as a thin well defined function appears to be very successful in representing low mass stars. Notice that the scatter in calculated temperatures is much narrower compared to the scatter in published temperatures for stars with $M\ll 1M_{\odot}$, as shown in Fig. 10. However, low mass stars in eclipsing binaries show a well known discrepancy of temperatures and radii with respect to the models, most likely related to the activity \citep*{Cakirli10, Morales10, Helminiak11, Bass12, Stassun13}. This may partially explain the larger spread in the low-mass regime for the published temperatures. Apparently, radii of the sample stars in the same low-mass regime were not affected as indicated on $M-R$ diagram (Fig. 7a), a smooth and narrow distribution is produced in the same low-mass region of stars in Fig. 10.   

Any well defined MLR provides a single $L$ for a given $M$. Metallicity and evolution information contained on the $M-L$ diagram, will be lost with this single value of $L$. This may appear as a draw back. On the contrary, this study asserts that the information lost by defining a MLR could be re-introduced into the $M-T_{eff}$ diagram by calculating effective temperatures using MLR and $R$. Accurately determined radii constrain the effects of metallicity and evolution because $R$ is one of the primary products of the evolution theory, which uses $M$ and metallicity as free initial parameters. One does not need to know age and chemical composition of the star because  it naturally propagates to $T_{eff}$ calculated. Since the same applies to all stars in the calibration sample, all calculated temperatures as calculated are homogeneous to reflect the evolution and metallicity information contained on $M-R$ diagram. In addition to this propagated effect of evolution, further thickening of the distribution for the stars $M>1M_{\odot}$ is introduced by errors. Therefore, the difference between ZAMS and TAMS reaches to 0.2 dex (in $\log T_{eff}$), rather than 0.15 dex implied by ZAMS and TAMS lines of \cite{Bertelli08, Bertelli09}.

\subsection{Applications with less accurate $M$ and $R$}
The presented method of calculating effective temperatures has been applied to a larger sample (371 stars) containing less accurate $M$ and $R$. The sample has been chosen from the 514 stars in the same catalogue from which the calibration sample was selected. For this larger sample there were only two selection rules: 1) both $M$ and $R$ could have errors up to 6\%, 2) both components had to be on the main sequence. Unlike for the calibration stars, there is no limitation on the accuracy of the luminosities here. There are 408 stars with $M$ and $R$ having errors less than or equal to 6\%. That number is reduced to 371 after removing non main-sequence stars. This new list naturally contains the calibration sample. The new calculated effective temperatures and published temperatures for these stars are listed in Table 7. Columns are self explanatory to indicate sequence number, name of the binary, equatorial coordinates, component (primary or secondary), relative errors in mass and radius, published temperature and its error, computed temperature and the upper limit of its error. Notice that in this new list there are 12 stars with no effective temperatures. New temperature estimates were calculated for these stars, and calculated temperatures for the entire sample are homogenized to a single method. Calculated temperatures have accuracies mostly better than 8\%. 

Figure 11a shows the $M-R$ diagram for the 371 main-sequence stars. Fig. 11b shows their distribution on the $M-L$ diagram. Finally, Fig. 11c illustrates the $M-T_{eff}$ diagram with effective temperatures computed. All temperatures are homogenized and despite having accuracies that are $\Delta T/T\leq 8$\%, stellar evolution within the main-sequence is also noticeable on the $M-T_{eff}$ diagram in the same fashion as in Fig. 10.

Figure 11 also demonstrates why main-sequence MLR is fundamentally confirmed and universally recognized, but the same is not true for the $M-R$ and $M-T_{eff}$ relations. If mass loss is neglected during the main-sequence lifetime, radius evolution pulls the star upward on the $M-R$ diagram, while corresponding temperature evolution is downward on the $M-T_{eff}$ diagram. This is at least what is happening in Fig. 11, where a single value of $L$ of a given $M$ is used according to derived MLR's in this study. A single value of $L$ means there is no evolutionary and metallicity effects on $L$. Evolutionary and metallicity effects are main contributing factors to the uncertainty of $L$. In fact, as discussed in previous sections, the observational uncertainties of $R$ and $T$, which  propagate to $L$, are negligible compared to evolutionary and metallicity effects. The upward evolutions seen for the stars $M>1M_{\odot}$ on $M-R$ diagrams become downward evolutions on the $M-T_{eff}$ diagram as a consequence of a fixed $L$. One should still keep in mind that the evolutionary and metallicity information contained in $M-R$ diagrams must also contained in $M-T_{eff}$ diagrams, albeit exaggerated by errors of the computed temperatures. 

If we look at stellar evolution theory in general, the stars with $M>1M_{\odot}$ the luminosity increases but in such a way that the surface temperature drops relatively little because of the expanding radius \citep{Clayton68} except in the final stages of MS evolution when the central convection zone starts to shrink and disappear by the depletion of hydrogen in the center. In this final stage, surface effective temperatures increase a little such that, the overall effect is that the star is brighter but cooler at TAMS than when it is at ZAMS. On the contrary, for the stars with $M<1M_{\odot}$, the temperature rises initially, to drop a little later while $L$ is continuously increasing from ZAMS to TAMS. Reversals of the theoretical ZAMS and TAMS lines of \citet{Bertelli08, Bertelli09} on Fig. 10 at about $M=1.05M_{\odot}$ are just indicating this fact. Notice that, for the stars with $M>1.05M_{\odot}$, the TAMS line indicates lower, but for the stars with $M<1.05M_{\odot}$ the TAMS line indicates higher temperature than the temperature at the ZAMS. 

Vertical spread of $T_{eff}$ in Fig. 11c is about one order of magnitude between coolest to hottest stars, while the vertical spread of radii is about two orders of magnitude in Fig. 11a. Consequently, the Stefan Boltzmann law requires $L$ to spread about 8 orders magnitudes between the limiting values of $L$ on Fig. 11b because L is proportional to the fourth power of the temperature and square of the radius. With such a large spread in vertical, the distribution of $L$ values is more suitable than $R$ and $T_{eff}$ distributions to be expressed by a single thin function, which is called MLR. Squeezing the three different scales on the figure into a similar paper scale, makes the thickness of $L$ smallest of all. The organization of stars in a sufficiently narrow band across the diagonal on the $M-L$ diagram clearly demonstrates the well-recognized universal MLR, and secret of the stellar realm discovered in the beginning of the 20th century. This study indicates that the universal MLR for stars in the solar neighborhood is not a continuous curve but rather a continuous sequence of four lines.

Absolute values of the differences between the calculated and published temperatures of Fig. 8 were divided by the published temperatures and analyzed as absolute relative differences, as shown in Fig. 12a. Most of the calculated temperatures agree with the published temperatures to within 10\%. Few cases show differences greater than 20\%. The calculated and published temperature errors of Fig. 8 are displayed as percentages in Fig. 12b and Fig. 12c, respectively. The relative errors of calculated temperatures appear as a nearly horizontal distribution because uncertainty contributions of $M$ and $R$ are negligible, and the four levels of dispersions shown are those based on the four MLRs given in Table 3. Distributions of $M$ and $R$ errors in \cite{Eker14} indicates the peak of the distribution is at 1\% and 2\% respectively, and for a smaller number of stars increases rapidly up to 5\% in both distributions. Therefore, allowing more stars with radius errors of up to 6\%, does not significantly change the  distribution of errors for calculated temperatures since the error contribution of the dispersions on the $M-L$ diagram dominates. 

\section{Conclusions}
A calibration sample has been formed to study $M-L$, $M-R$ and $M-T_{eff}$ diagrams of nearby main-sequence stars. The stars were selected from Table 2 of \cite{Eker14} with three basic conditions: 1) Stars must be on the main-sequence, 2) masses and radii relative errors must be less than or equal to 3\%, and 3) luminosity errors must be less than or  equal to 30\%. Being components of detached double-lined eclipsing binaries, the sample stars have not yet experienced any mass transfer. All can be taken as evolved as single stars.  

The efficiency of stellar furnaces, that is $\log (L/M)$, is found to increase linearly from 0.38 to $1.05M_{\odot}$. The quantity $\log (L/M)$ continues increasing, but with a smaller slope from $1.05M_{\odot}$ to $2.4M_{\odot}$. We have found three break points, dividing the present sample into four subsamples, which we identified as stellar mass domains: low mass ($0.38<M/M_{\odot}\leq 1.05$), intermediate mass ($1.05<M/M_{\odot}\leq 2.4$), high mass ($2.4<M/M_{\odot}\leq 7$), and very high mass ($M/M_{\odot}>7$). 

Those stellar mass domains were used to revise the classical MLR ($L \propto M^{\alpha}$). Lines with different $\alpha$ were fit to the masses and luminosities in each domain and best fitting lines were determined by the least squares method. The quality of all fits is very similar, but the four-piece linear function stands out with a physical background. The $M-R$ and $M-T_{eff}$ diagrams of the present sample have also been studied. Stellar evolution within the main-sequence band is clearly apparent in the $M-R$ diagram. Corresponding evolution could not be seen however, on the $M-T_{eff}$ diagram based on published temperatures.

A well known method of calculating stellar effective temperatures for main-sequence stars has been discussed and analyzed. The calculated temperatures should have accuracies of $\pm$8\% if $M$ and $R$ errors are less than or equal to 6\%. The method is still applicable to less accurate stars. With less accurate stars, one must propagate observational errors of the mass to the predicted luminosity. The method produces correspondingly less accurate temperatures with increasingly less accurate mass and radius as inputs.

The method has been applied to a calibration sample and a wider sample containing 12 stars without published temperatures. The main-sequence evolution, which was clear on the $M-R$ diagram, but not seen on the $M-T_{eff}$ diagram with published temperatures, becomes clearly visible with the empirical effective temperatures calculated. Stellar temperatures based on this method are real effective temperatures, obtained directly form absolute stellar properties ($M$, $R$). In comparison, published temperatures are based on apparent properties, including colors or spectral lines, or through Pogson's formula using parallax and apparent magnitude. Alternative temperature estimates therefore, as previously published, require bolometric corrections. We believe that stellar temperatures calculated by the method based on the detached double-lined eclipsing binaries in the present study have relatively larger errors than published ones, but have potential in calibrating bolometric corrections needed to estimate the stellar temperatures of main sequence stars in general, including in single, multiple, and other kinds of binary systems.

\section{Acknowledgments}
Authors are grateful to the anonymous referee whose comments were very 
useful in improving the manuscript. This work has been supported in 
part by the Scientific and Technological Research Council (T\"UB\.ITAK) grant 
numbers 106T688 and 111T224. This research has made use of the SIMBAD 
database, operated at CDS, Strasbourg, France and NASA's Astrophysics 
Data System Bibliographic Services.

\pagebreak

\begin{figure*}
\begin{center}
\includegraphics[scale=0.80, angle=0]{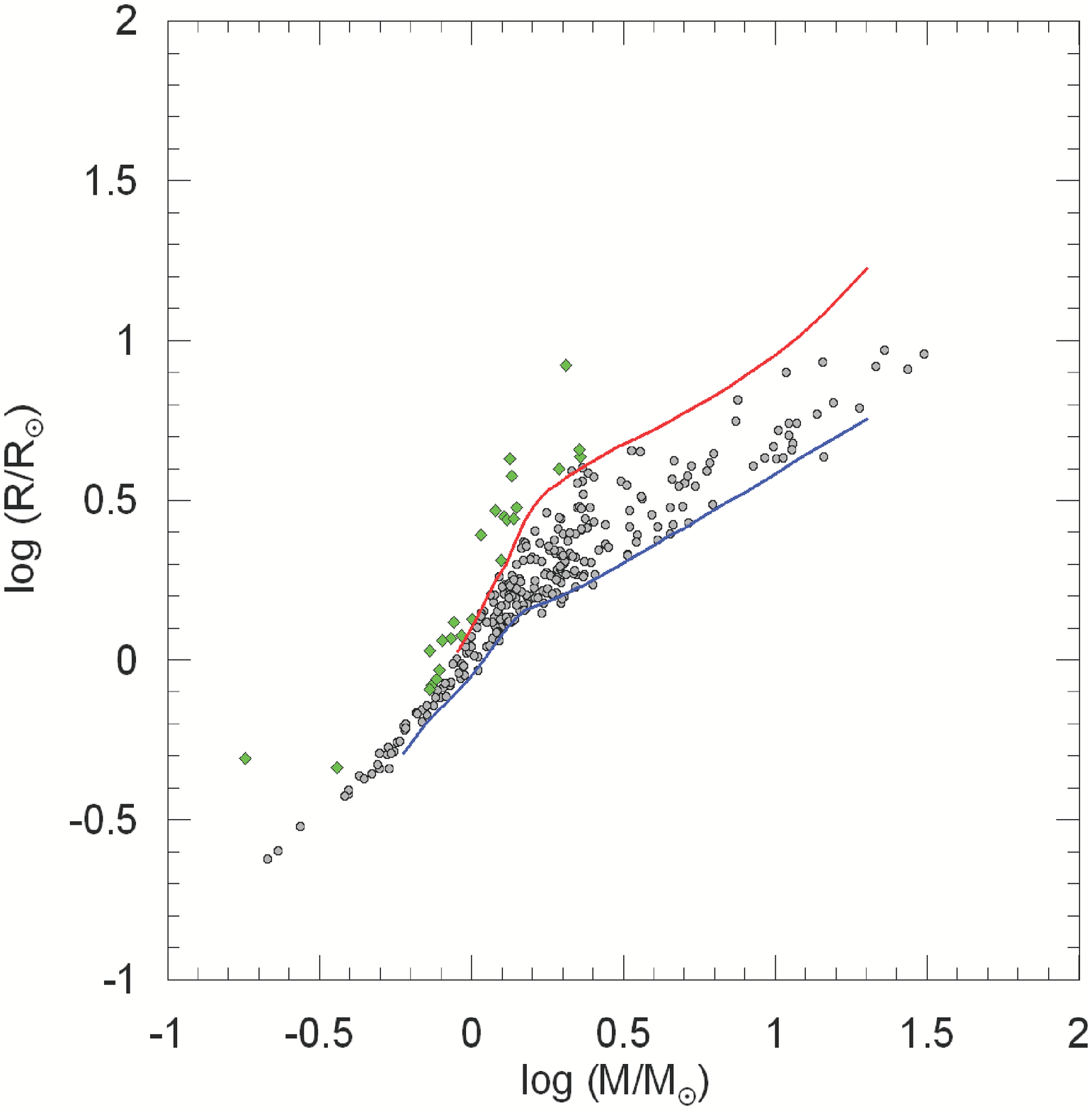}
\caption[] {The masses and radii of main-sequence stars (filled circles) in the calibration sample is secured according to their positions in between ZAMS and TAMS lines for zero metallicity from \citet{Bertelli08, Bertelli09}. The stars above TAMS line (diamonds) were designated to be probable non-main-sequence stars.}
\end{center}
\end{figure*}

\begin{figure*}
\begin{center}
\includegraphics[scale=0.80, angle=0]{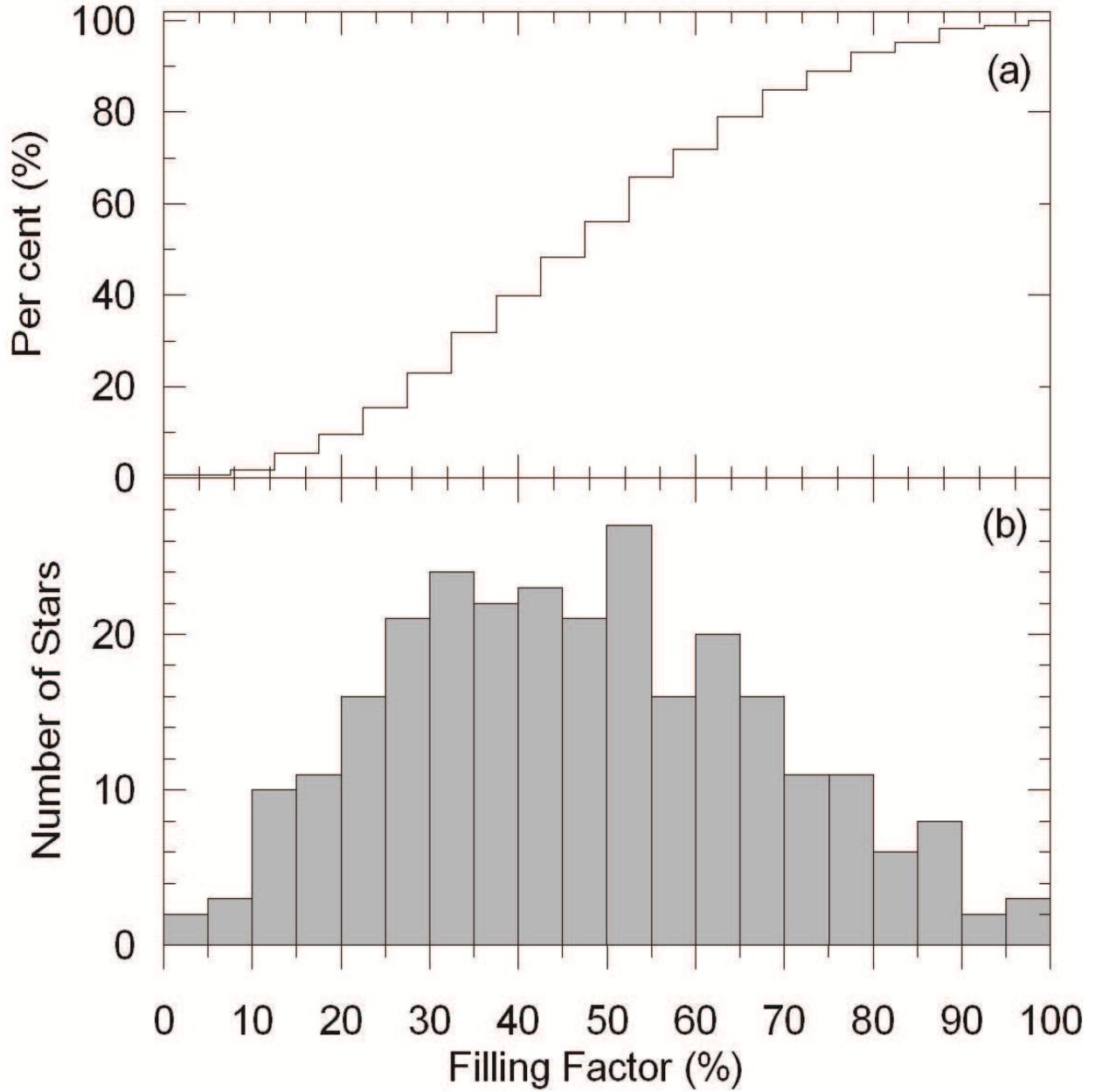}
\caption[] {Cumulative (a) and histogram (b) distribution of the filling factors for the stars in the calibration sample. Stars are spherical within 1\% of radius up to a filling factor of 75\% \citep{Eker14}.}
\end{center}
\end{figure*}

\pagebreak

\begin{figure*}
\begin{center}
\includegraphics[scale=0.70, angle=0]{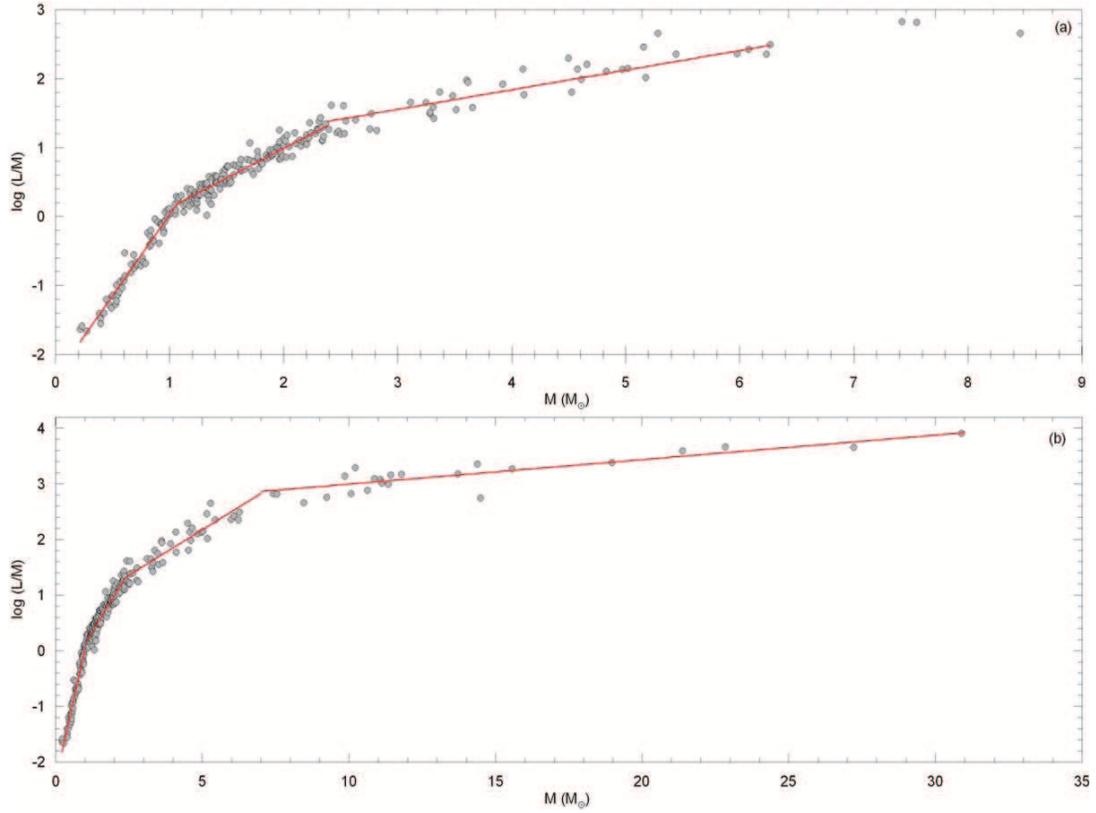}
\caption[] {Energy production rate per stellar mass ($L/M$). Separate linear distributions in $\log (L/M)$ versus $M$ are clear. Break points are at 1) $M=1.05M_{\odot}$, 2) $M=2.4M_{\odot}$, and 3) $M=7M_{\odot}$.}
\end{center}
\end{figure*}

\begin{figure*}
\begin{center}
\includegraphics[scale=0.70, angle=0]{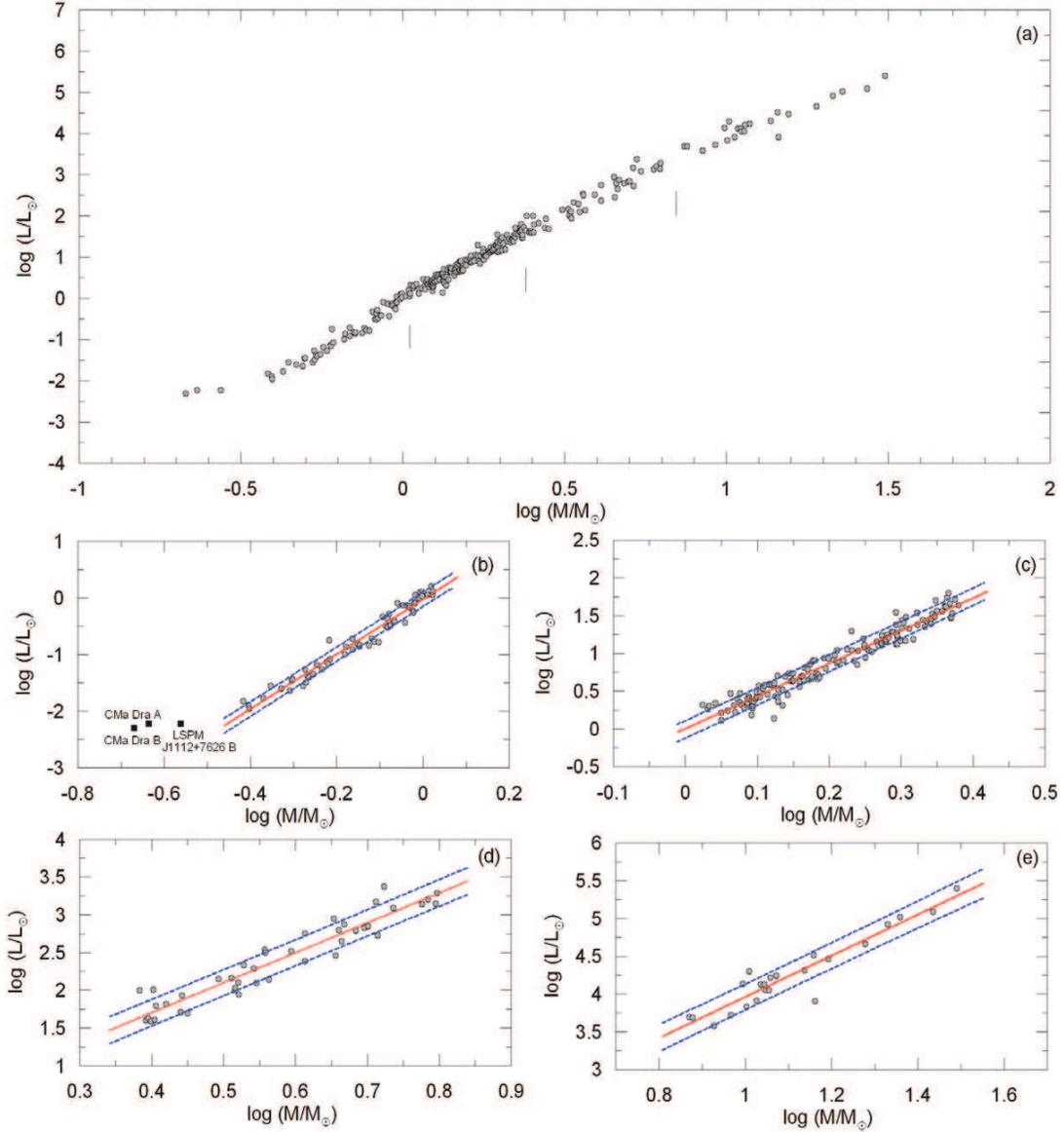}
\caption[] {The $M-L$ diagram of the calibration sample (a). The three vertical dashes corresponding to the break points in Fig. 3 mark low mass ($0.2<M/M_{\odot}\leq1$), intermediate mass ($1.05<M/M_{\odot}\leq2.4$), high mass ($2.4<M/M_{\odot}\leq7$) and very high mass ($7<M/M_{\odot}<32$) domains. The lower four panels (b, c, d, e): show best fitting lines and $1\sigma$ limits in those domains. Note: lower three points indicated for low mass domain were excluded from the analysis.}
\end{center}
\end{figure*}

\begin{figure*}
\begin{center}
\includegraphics[scale=0.70, angle=0]{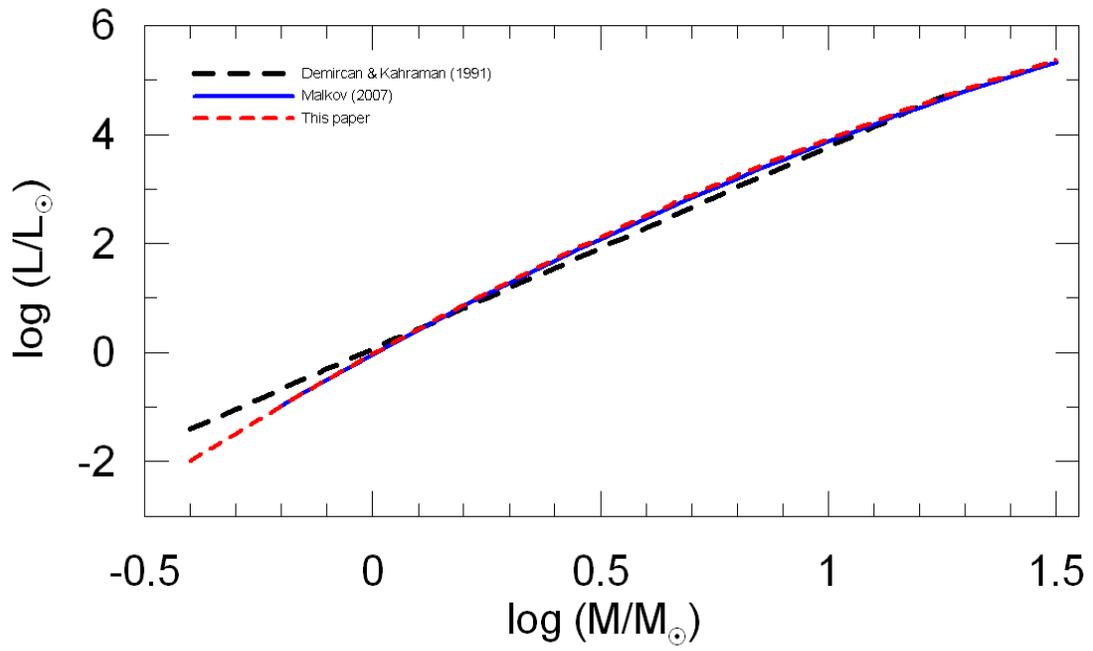}
\caption[] {A best fitting quadratic curve as the MLR for the whole sample compared to the quadratic MLRs of \cite{Malkov07} and \cite{Demircan91}.}
\end{center}
\end{figure*}

\begin{figure*}
\begin{center}
\includegraphics[scale=0.70, angle=0]{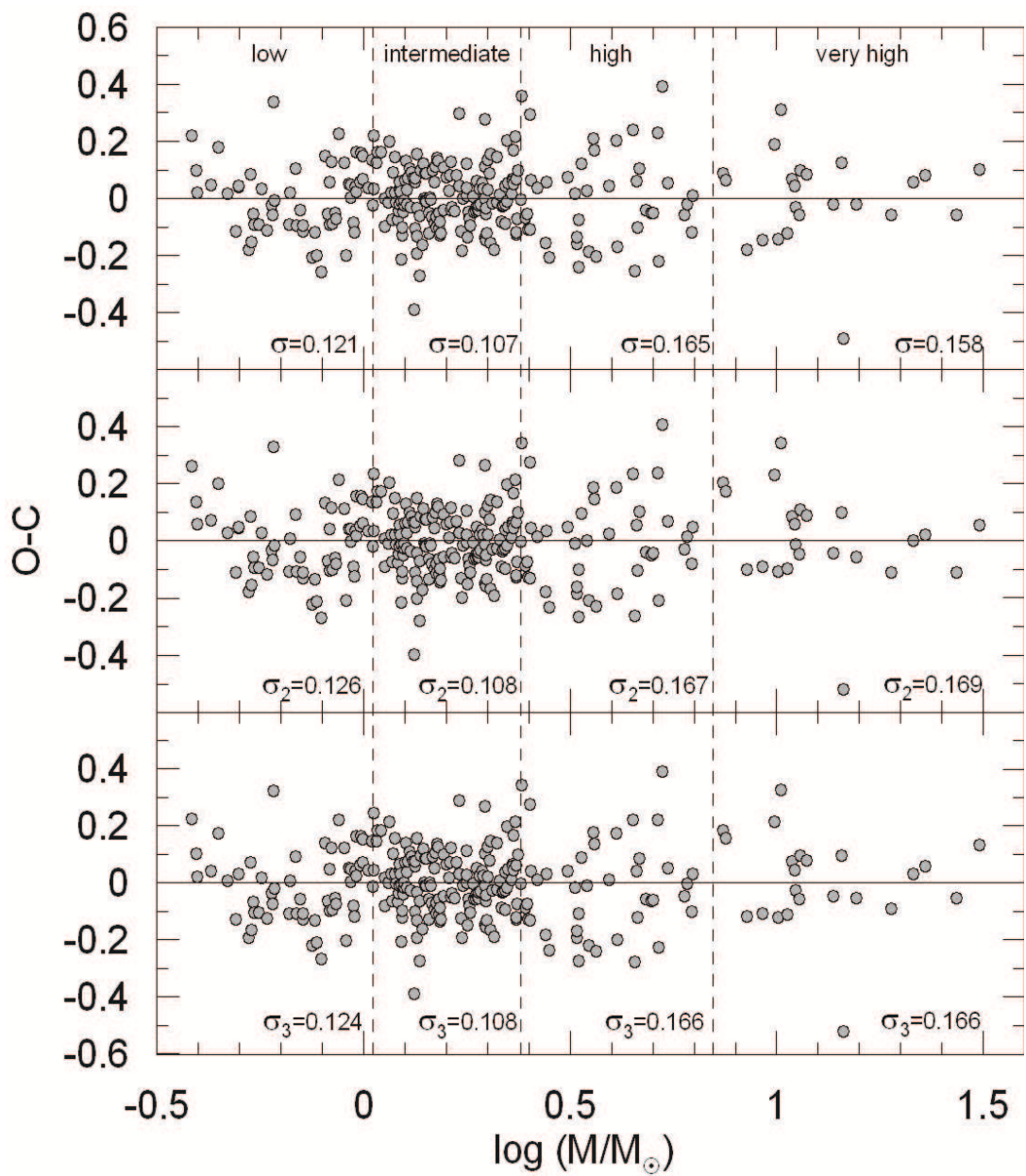}
\caption[] {The $O-C$ dispersions of the four piece linear MRLs (upper) compared to quadratic (middle) and cubic  (lower) MRLs. Low, intermediate, high and very high mass intervals separated by vertical dashed lines. Standard deviations of each segment ($\sigma$) indicates that the four piece linear MRLs (upper) are best to represent $M-L$ diagram in Fig. 4.}
\end{center}
\end{figure*}

\begin{figure*}
\begin{center}
\includegraphics[scale=0.70, angle=0]{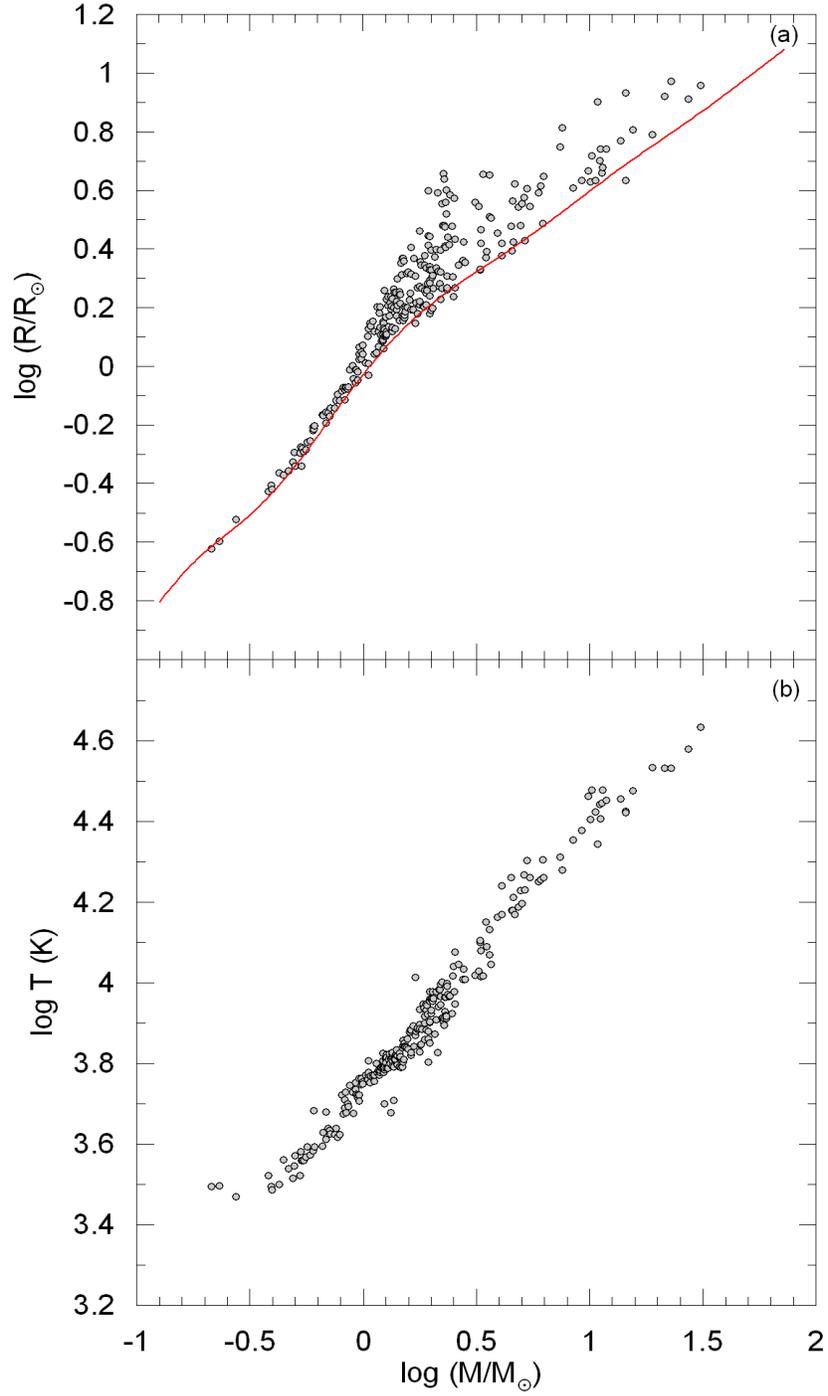}
\caption[] {Observational a) radii and b) effective temperatures versus mass. The continuous line on the mass-radius plot is theoretical ZAMS of \citet{Bertelli08, Bertelli09}.}
\end{center}
\end{figure*}

\begin{figure*}
\begin{center}
\includegraphics[scale=0.60, angle=0]{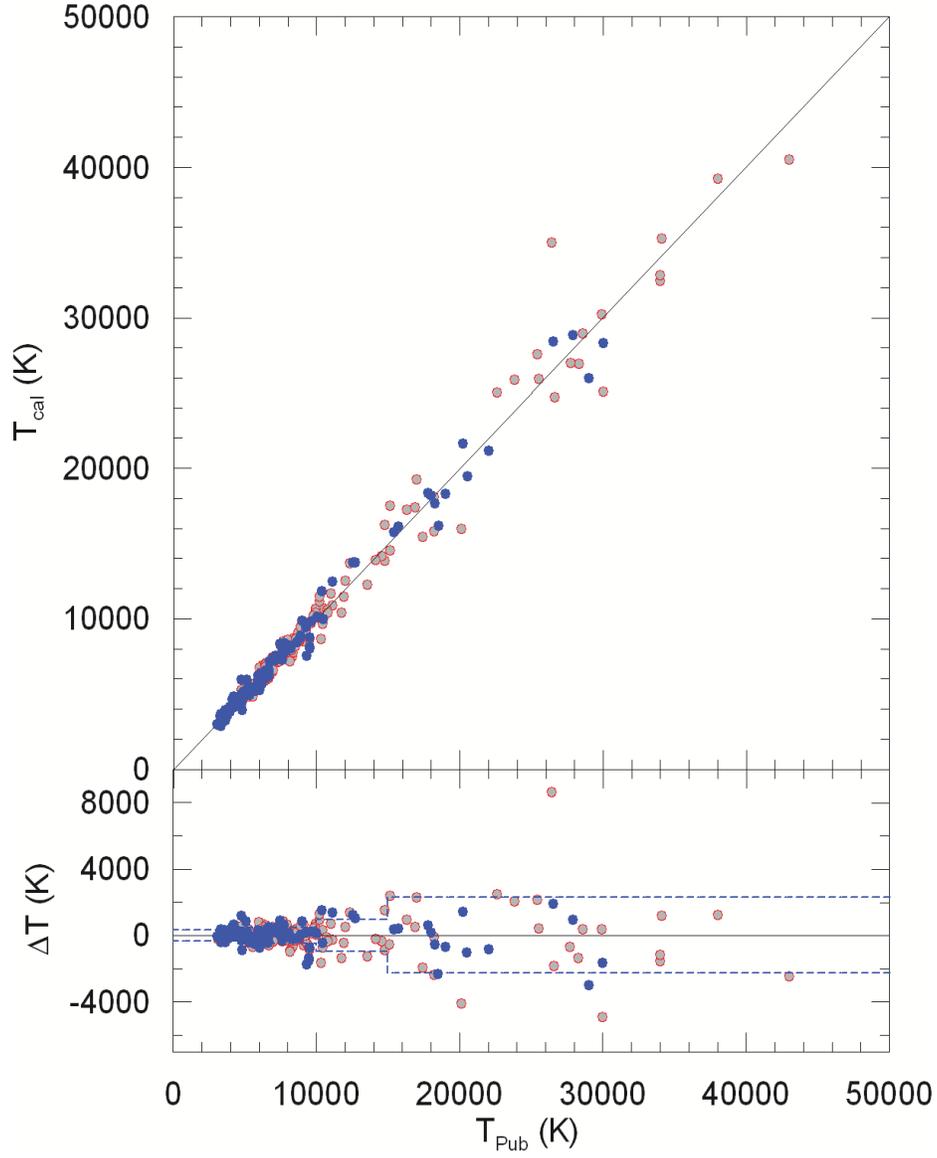}
\caption[] {Comparing the calculated (empirical) $T_{eff}$ and published $T_{eff}$ temperatures (above) and ($\Delta T = T_{cal}-T_{pub}$) the difference between them (below). Dashed: the mean standard differences ($\sqrt{\langle(T_{cal}-T_{pub})^2\rangle}$), the filled and empty circles: temperatures published within the last seven years and those published earlier, respectively.}
\end{center}
\end{figure*}

\begin{figure*}
\begin{center}
\includegraphics[scale=0.60, angle=0]{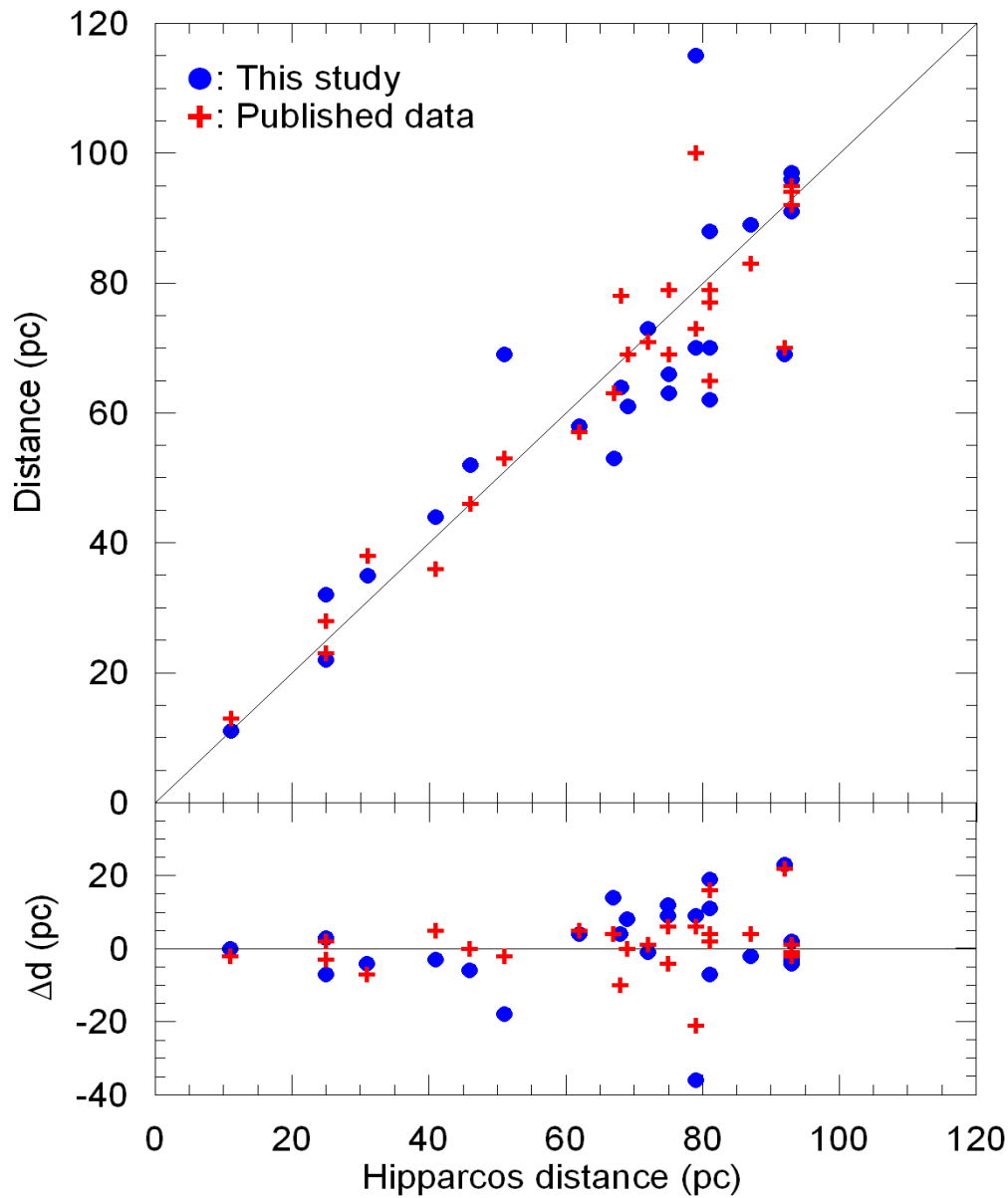}
\caption[] {Photometric distances of some selected stars ($\sigma_{\pi}/\pi<0.15$) in the calibration sample were compared to $Hipparcos$ distances.  Photometric distances were estimated as explained in the text using calculated temperatures (circles) and published temperatures (plus). Photometric and $Hipparcos$ distances (above) and ($\Delta d = d_{Hip}-d_{Pho}$) the difference from the $Hipparcos$ distances (below).}
\end{center}
\end{figure*}

\begin{figure*}
\begin{center}
\includegraphics[scale=0.60, angle=0]{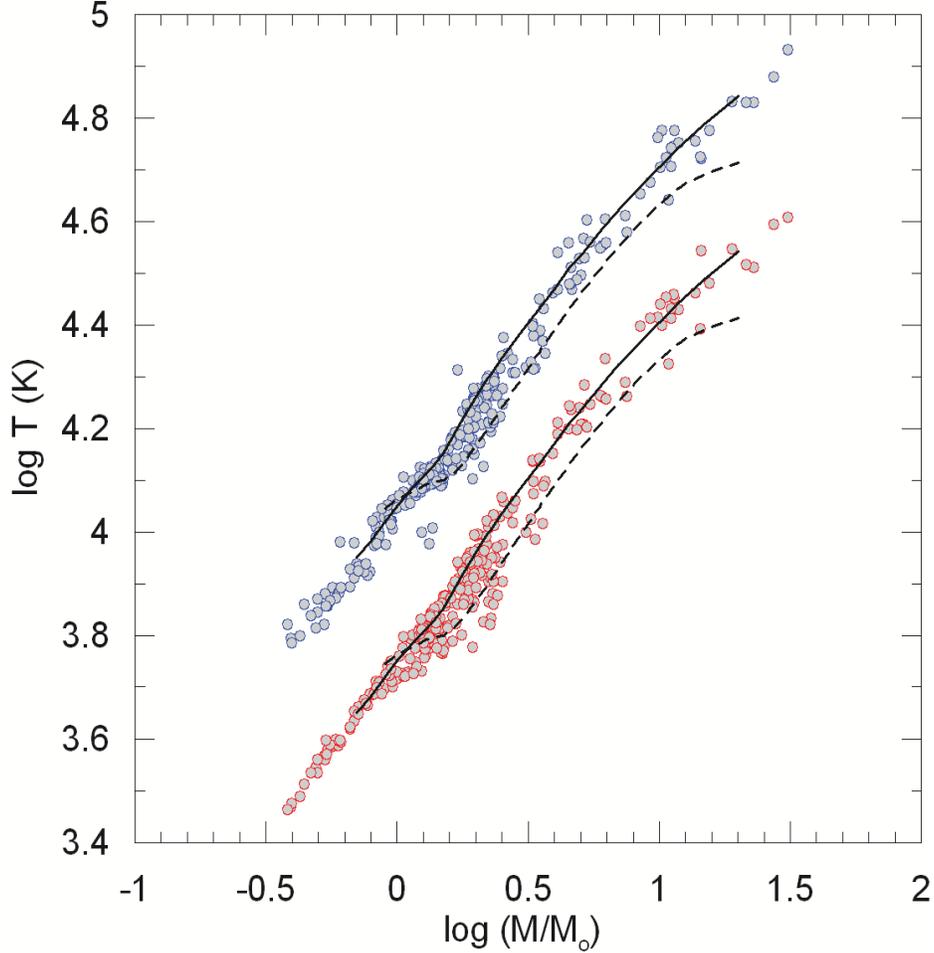}
\caption[] {$M-T_{eff}$ diagrams. The upper shows published temperatures, as shown in Fig. 7b, but shifted up in vertical scale by +0.3 dex for side-by-side comparison. The lower shows calculated temperatures with empirical $T_{eff}$ calculated according to Eq. (1). ZAMS (solid) and TAMS (dashed) lines of \cite{Bertelli08, Bertelli09} indicates theoretical thicknesses of $M-T_{eff}$ diagram for main-sequence stars. Larger errors of calculated temperatures makes main-sequence thicker, while a thin MRL better represents main-sequence stars with $M<1M_{\odot}$, so the large degree of scattering (upper) for stars with $M<1M_{\odot}$.}
\end{center}
\end{figure*}

\begin{figure*}
\begin{center}
\includegraphics[scale=0.80, angle=0]{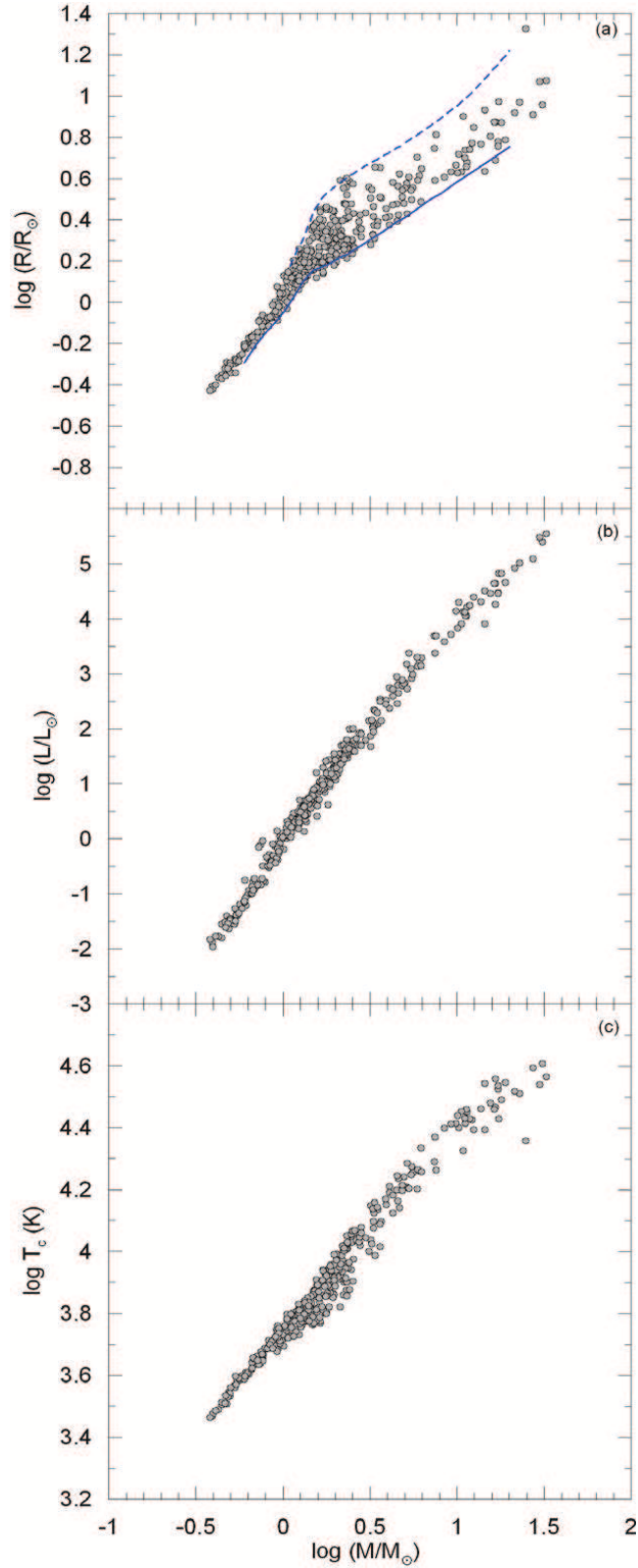}
\caption[] {a) Larger sample ($N=371$) of main-sequence stars on $M-R$ diagram. b) Their $M-L$ diagram (excluding 12 without published $T_{eff}$) and c) their $M-T_{eff}$ diagram with calculated effective temperatures. The solid lines and dashed are theoretical ZAMS and TAMS from \citet{Bertelli08, Bertelli09}.}
\end{center}
\end{figure*}

\begin{figure*}
\begin{center}
\includegraphics[scale=0.80, angle=0]{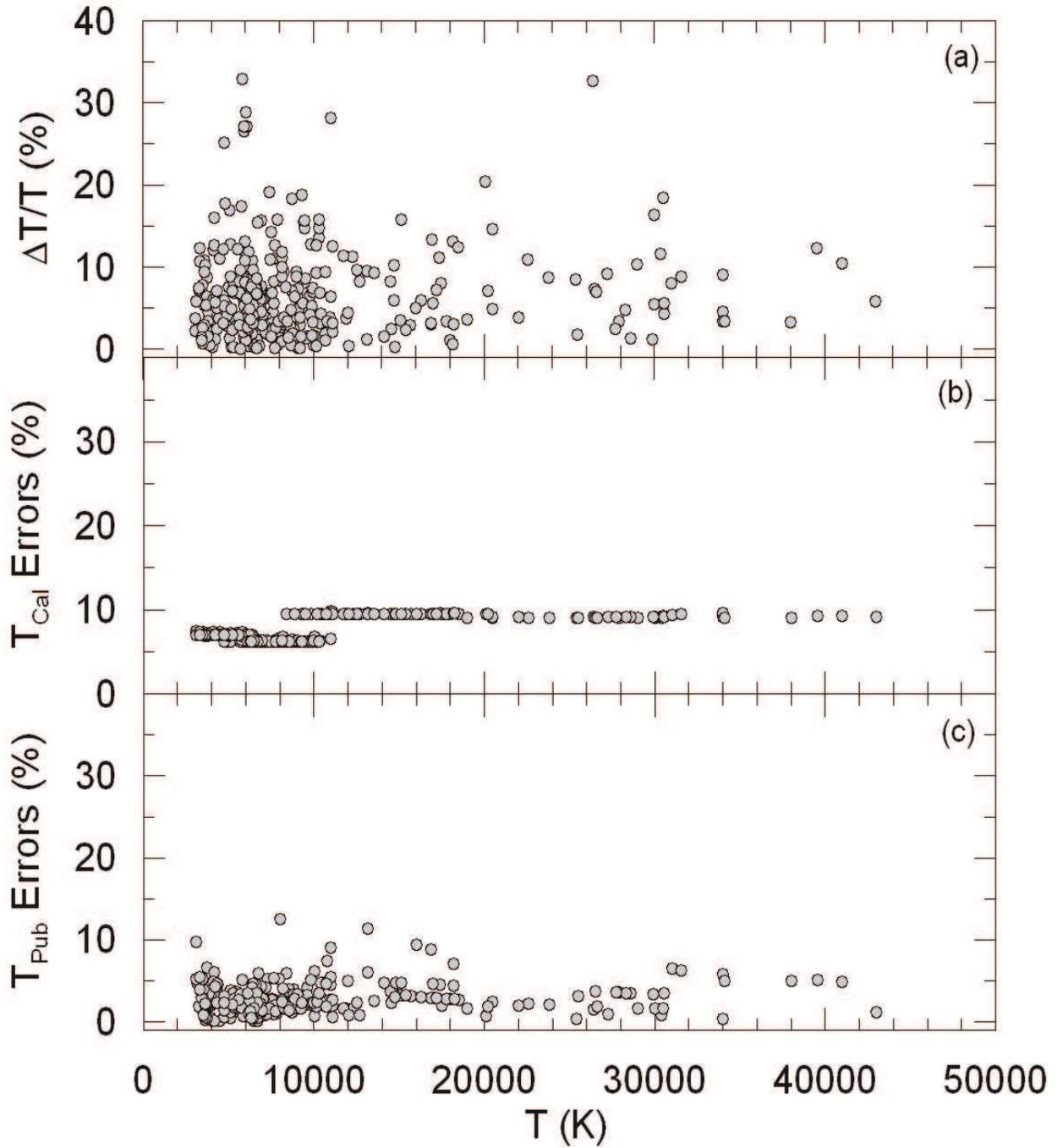}
\caption[] {a) Comparing absolute relative differences between calculated and published temperatures of larger sample. b) Relative errors of calculated and, c) relative errors of published temperatures.}
\end{center}
\end{figure*}

\pagebreak
\pagebreak

\begin{table*}[h] 
\setlength{\tabcolsep}{.3pt}
\begin{center}
\caption{Comparison of data samples for earlier MLRs and MRRs.}
\scriptsize{
\begin{tabular}{|c|c|c|c|c|c|}
\hline
\multicolumn{2}{|c|}{Sample} &   Accuracy  &    Mass-Luminosity        &  Mass-Radius          & Ranges \\
\hline
This Study           &268 stars (all components of detached  & $M \&R \leq 3\%$     & Four broken lines,       &              ---  &$0.2<M/M_{\odot}<32$     \\
                     &double-lined eclipsing), homogenized   & $L \leq 30\%$        & linear and quadratic     &                   &$0.23<R/R_{\odot}<9.36$  \\
\hline
Torres et al. (2010) &190 stars (94 eclipsing+$\alpha$ Cen)  & $M \&R \leq 3\%$     & $\log M$-$\log L$        & $\log M$-$\log R$ &$0.21<M/M_{\odot}<27.27$ \\
                     &homogenized                            &                      & diagram displayed        & diagram displayed &$0.24<R/R_{\odot}<9.35$  \\
\hline
Malkov (2007)        &215 stars (114 detached eclipsing),    & $M \&R \leq 10\%$    & Quadratic                & Cubic             & $0.63<M/M_{\odot}<31.6$ \\
                     &as published                           &                      & $\sigma=0.12$            & $\sigma=0.08$     & $0.63<R/R_{\odot}<25.1$ \\
\hline
Henry (2004)         &188 stars (139 detached main-sequence, & $M \&R \leq 15\%$    & Mass-$M_V$, plot         &                   &$0.07<M/M_{\odot}<33$    \\
\hline
\end{tabular}
}
\end{center}
\end{table*}

\begin{table*}
\setlength{\tabcolsep}{1pt}
\renewcommand{\arraystretch}{.7}
\caption{The basic stellar parameters and relative errors for the stars selected for the calibration sample.}
\tiny{
\begin{tabular}{rlrrrcccccccccc}
\hline
  ID & Star name &  $\alpha$(J2000) & $\delta$(J2000) &  Com & $M/M_{\odot}$ & $\Delta M_{err}/M$ &    $R/R_{\odot}$ & $\Delta R_{err}/R$ &  $T_{eff}$ & $(T_{eff})_{err}$ & $\log L/L_{\odot}$ & $\Delta L_{err}/L$ &FF($\%$) & Remark\\
\hline
			1     & MU Cas & 00:15:51.56 & +60:25:53.64 & Pri   & 4.657 & 0.021 & 4.192 & 0.012 & 14750 & 500   & 2.873 & 0.056 & 37    & $\surd$ \\
			2     & MU Cas & 00:15:51.56 & +60:25:53.64 & Sec   & 4.575 & 0.020 & 3.671 & 0.011 & 15100 & 500   & 2.799 & 0.055 & 33    & $\surd$ \\
			3     & V69-47 Tuc  & 00:22:52.95 & -72:03:40.68 & Pri   & 0.876 & 0.006 & 1.316 & 0.004 & 5945  & 150   & 0.288 & 0.042 & 11    & X \\
			4     & V69-47 Tuc  & 00:22:52.95 & -72:03:40.68 & Sec   & 0.859 & 0.007 & 1.163 & 0.005 & 5959  & 150   & 0.185 & 0.042 & 9     & X \\
			5     & YZ Cas & 00:45:39.08 & +74:59:17.06 & Pri   & 2.308 & 0.004 & 2.547 & 0.012 & 9200  & 300   & 1.620 & 0.054 & 41    & $\surd$ \\
			6     & YZ Cas & 00:45:39.08 & +74:59:17.06 & Sec   & 1.347 & 0.007 & 1.359 & 0.015 & 6700  & 250   & 0.524 & 0.061 & 35    & $\surd$ \\
			7     & NGC188 KR V12 & 00:52:37.54 & +85:10:34.65 & Pri   & 1.102 & 0.006 & 1.425 & 0.013 & 5900  & 100   & 0.344 & 0.031 & 27    & $\surd$ \\
\hline
		\end{tabular}
		(X) Probable non main-squence star  (locates above ZAMS line in Fig. 1).
	}
\end{table*}


\begin{table*}
	\setcounter{table}{1} 
	\setlength{\tabcolsep}{1pt}
	\renewcommand{\arraystretch}{.7}
	\tiny{
		\begin{tabular}{rlrrrcccccccccc}
			\hline
			ID & Star name &  $\alpha$(J2000) & $\delta$(J2000) &  Com & $M/M_{\odot}$ & $\Delta M_{err}/M$ &    $R/R_{\odot}$ & $\Delta R_{err}/R$ &  $T_{eff}$ & $(T_{eff})_{err}$ & $\log L/L_{\odot}$ & $\Delta L_{err}/L$ &FF($\%$) & Remark\\
			\hline
			76    & RR Lyn & 06:26:25.84 & +56:17:06.35 & Pri   & 1.935 & 0.004 & 2.579 & 0.008 & 7570  & 100   & 1.292 & 0.023 & 29    & $\surd$ \\
			77    & RR Lyn & 06:26:25.84 & +56:17:06.35 & Sec   & 1.520 & 0.003 & 1.596 & 0.019 & 6980  & 100   & 0.735 & 0.029 & 23    & $\surd$ \\
			78    & V578 Mon & 06:32:00.61 & +04:52:40.90 & Pri   & 10.212 & 0.005 & 5.228 & 0.011 & 30000 & 500   & 4.298 & 0.030 & 66    & $\surd$ \\
			79    & V578 Mon & 06:32:00.61 & +04:52:40.90 & Sec   & 14.482 & 0.006 & 4.318 & 0.016 & 26400 & 400   & 3.910 & 0.029 & 68    & $\surd$ \\
			80    & WW Aur & 06:32:27.18 & +32:27:17.63 & Pri   & 1.964 & 0.004 & 1.928 & 0.006 & 7960  & 420   & 1.127 & 0.083 & 51    & $\surd$ \\
			81    & WW Aur & 06:32:27.18 & +32:27:17.63 & Sec   & 1.814 & 0.004 & 1.842 & 0.006 & 7670  & 410   & 1.023 & 0.084 & 52    & $\surd$ \\
			82    & GX Gem & 06:46:09.13 & +34:24:52.83 & Pri   & 1.488 & 0.007 & 2.334 & 0.005 & 6194  & 100   & 0.857 & 0.027 & 51    & $\surd$ \\
\hline
		\end{tabular}
		(X) Probable non main-squence star (locates above ZAMS line in Fig. 1).
	}
\end{table*}

\begin{table*}
	\setlength{\tabcolsep}{1pt}
	\renewcommand{\arraystretch}{.7}
	\tiny{
		\begin{tabular}{rlrrrcccccccccc}
			\hline
			ID & Star name &  $\alpha$(J2000) & $\delta$(J2000) &  Com & $M/M_{\odot}$ & $\Delta M_{err}/M$ &    $R/R_{\odot}$ & $\Delta R_{err}/R$ &  $T_{eff}$ & $(T_{eff})_{err}$ & $\log L/L_{\odot}$ & $\Delta L_{err}/L$ &FF($\%$) & Remark\\
			\hline
			151   & IM Vir & 12:49:38.70 & -06:04:44.86 & Sec   & 0.664 & 0.008 & 0.681 & 0.019 & 4250  & 130   & -0.866 & 0.051 & 46    & $\surd$ \\
			152   & eta Mus & 13:15:14.94 & -67:53:40.52 & Pri   & 3.283 & 0.012 & 2.140 & 0.009 & 12700 & 100   & 2.029 & 0.016 & 51    & $\surd$ \\
			153   & eta Mus & 13:15:14.94 & -67:53:40.52 & Sec   & 3.282 & 0.012 & 2.130 & 0.019 & 12550 & 300   & 2.005 & 0.043 & 51    & $\surd$ \\
			154   & SZ Cen & 13:50:35.09 & -58:29:57.11 & Pri   & 2.272 & 0.011 & 4.555 & 0.005 & 8000  & 300   & 1.883 & 0.061 & 75    & X \\
			155   & SZ Cen & 13:50:35.09 & -58:29:57.11 & Sec   & 2.311 & 0.009 & 3.622 & 0.006 & 8280  & 300   & 1.743 & 0.059 & 64    & $\surd$ \\
			156   & DM Vir & 14:07:52.43 & -11:09:07.49 & Pri   & 1.454 & 0.006 & 1.764 & 0.010 & 6500  & 100   & 0.698 & 0.027 & 37    & $\surd$ \\
			157   & DM Vir & 14:07:52.43 & -11:09:07.49 & Sec   & 1.448 & 0.006 & 1.764 & 0.010 & 6500  & 300   & 0.698 & 0.074 & 37    & $\surd$ \\
\hline
		\end{tabular}
		\\
		(X) Probable non main-squence star (locates above ZAMS line in Fig. 1).
	}
\end{table*}

\begin{table*}
	\setlength{\tabcolsep}{1pt}
	\renewcommand{\arraystretch}{.7}
	\tiny{
		\begin{tabular}{rlrrrcccccccccc}
			\hline
			ID & Star name &  $\alpha$(J2000) & $\delta$(J2000) &  Com & $M/M_{\odot}$ & $\Delta M_{err}/M$ &    $R/R_{\odot}$ & $\Delta R_{err}/R$ &  $T_{eff}$ & $(T_{eff})_{err}$ & $\log L/L_{\odot}$ & $\Delta L_{err}/L$ &FF($\%$) & Remark\\
			\hline
			226   & V1430 Aql & 19:21:48.49 & +04:32:56.92 & Pri   & 0.957 & 0.010 & 1.101 & 0.009 & 5262  & 150   & -0.079 & 0.047 & 70    & $\surd$ \\
			227   & V1430 Aql & 19:21:48.49 & +04:32:56.92 & Sec   & 0.859 & 0.023 & 0.851 & 0.012 & 4930  & 100   & -0.416 & 0.035 & 60    & $\surd$ \\
			228   & UZ Dra & 19:25:55.05 & +68:56:07.15 & Pri   & 1.345 & 0.015 & 1.311 & 0.023 & 6210  & 110   & 0.361 & 0.035 & 35    & $\surd$ \\
			229   & UZ Dra & 19:25:55.05 & +68:56:07.15 & Sec   & 1.236 & 0.016 & 1.150 & 0.017 & 5985  & 110   & 0.183 & 0.034 & 34    & $\surd$ \\
			230   & V2080 Cyg & 19:26:47.95 & +50:08:43.77 & Pri   & 1.190 & 0.014 & 1.596 & 0.005 & 6000  & 75    & 0.472 & 0.022 & 35    & $\surd$ \\
			231   & V2080 Cyg & 19:26:47.95 & +50:08:43.77 & Sec   & 1.156 & 0.015 & 1.599 & 0.005 & 5987  & 75    & 0.470 & 0.022 & 36    & $\surd$ \\
			232   & V885 Cyg & 19:32:49.86 & +30:01:17.03 & Pri   & 2.000 & 0.015 & 2.487 & 0.005 & 8375  & 150   & 1.436 & 0.030 & 64    & $\surd$ \\
\hline
		\end{tabular}
		(X) Probable non main-squence star (locates above ZAMS line in Fig. 1).
	}
\end{table*}

\begin{table*}
\setcounter{table}{2} 
\setlength{\tabcolsep}{2pt}
{\small
\begin{center}
\caption{Classical MLRs for the four mass domains.}
\begin{tabular}{rcccccc}
\hline
Domain            &$N$ & Mass range             &   Equation                              &   $R$ &  $\sigma$ &   $\alpha$ \\
\hline
Low mass          & 57 & $0.38<M/M_{\odot}\leq1.05$ & $\log L=4.841(132)\times \log M- 0.026(25)$ & 0.980 & 0.121 &  4.841 \\
Intermediate mass &146 & $1.05<M/M_{\odot}\leq2.40$ & $\log L=4.328(90) \times \log M-0.002(20)$ & 0.970 & 0.108 &  4.328 \\
High mass         & 42 & $2.4<M/M_{\odot}\leq7$     & $\log L=3.962(203)\times \log M+0.120(112)$& 0.951 & 0.165 &  3.962 \\
Very high mass    & 23 & $7<M/M_{\odot}<32$         & $\log L=2.726(203)\times \log M+1.237(228)$& 0.946 & 0.158 &  2.726 \\
\hline
\end{tabular}  
\end{center}
}
\end{table*}

\begin{table*}
\setlength{\tabcolsep}{2pt}
{\small
\begin{center}
\caption{Comparing linear, quadratic and cubic MLRs of the calibration sample as a whole (268 main-sequence stars).}   
\begin{tabular}{lcc}
\hline
Equation          &  $R$ &  $\sigma$\\
\hline
$\log L=+4.040(32)\times (\log M)-0.002(14)$ & 0.992 & 0.187\\
$\log L=-0.705(41)\times(\log M)^2+4.655(42)\times (\log M)-0.025(10)$ & 0.992 & 0.130\\
$\log L=-0.125(84)\times(\log M)^3-0.535(121)\times(\log M)^2+4.634(45)\times (\log M)-0.033(11)$ & 0.992 & 0.130\\
\hline
\end{tabular}  
\end{center}
}
\end{table*}

\begin{table*}
\setlength{\tabcolsep}{2pt}
{\small
\begin{center}
\caption{Standard deviations on M-L diagram and corresponding relative uncertainties.}
\begin{tabular}{lcccccccc}
\hline
Domain            & Mass range &  $\sigma$ & $\Delta L/L (\%)$ & $(\Delta L/L)/4$ & $(\Delta R/R)/2$ & $\Delta T/T$ & $\alpha$ & $\Delta M/M(\%)$ \\
\hline
Low mass          & $0.38<M/M_{\odot}\leq1.05$  &  0.121 & 27.86 & 6.96 & 1.5 & 7.12 & 4.841 & 5.76 \\
Intermediate mass & $1.05<M/M_{\odot}\leq2.40$  &  0.108 & 24.87 & 6.22 & 1.5 & 6.39 & 4.328 & 5.75 \\
High mass         & $2.40<M/M_{\odot}\leq7.00$  &  0.165 & 37.99 & 9.50 & 1.5 & 9.62 & 3.962 & 9.59 \\
Very high mass    & $7<M/M_{\odot}<32$          &  0.158 & 36.38 & 9.10 & 1.5 & 9.22 & 2.726 & 13.34\\
\hline
\end{tabular}  
\end{center}
}
\end{table*}

\begin{table*}
\setcounter{table}{5} 
\setlength{\tabcolsep}{5pt}
{\small
\begin{center}
\caption{Comparing the calculated and the published temperatures.}
\begin{tabular}{rcccc}
\hline
$T_{eff}$ range (K) & $N$ & Mean standard & Mean calculated & Mean published \\
                    &     &  difference (K)     & error  (K) & error  (K)\\
\hline
(2750, 5000]     &  37 &  345 &  291 & 117 \\
(5000, 10000]    & 168 &  441 &  453 & 162 \\
(10000, 15000]   &  24 &  975 & 1109 & 392 \\
(15000, 43000]   &  39 & 2283 & 2203 & 746 \\
\hline
\end{tabular}  
\end{center}
Mean standard difference=$\sqrt{\frac{\Sigma (T_{cal}-T_{pub})^2}{N}}$. Summation over all stars in the range given.\\
}
\end{table*}  

\begin{table*}
	\setlength{\tabcolsep}{1pt}
 \renewcommand{\arraystretch}{.7}
 \caption{Comparing published and calculated (empirical) effective temperatures.}
 \tiny{
  \begin{tabular}{clcccccccclccc}
   \hline
   &    &     &     &     &      &      & \multicolumn{4}{c}{Published temperatures}    & \multicolumn{3}{c}{This study}\\
   \hline
   ID &       Star &   $\alpha$(J2000) & $\delta$(J2000) & Com & $\Delta M/M$ & $\Delta R/R$ &    $T_{eff}$ & $\Delta T_{eff}$ & $\Delta T/T$ & \multicolumn{1}{c}{Ref.} &    $T_{eff}$ & $\Delta T_{eff}$ & $\Delta T/T$ \\
   \hline
   1 & DV Psc & 00:13:09.20 & +05:35:43.06 & Pri & 0.028 & 0.044 & 4450 & $-$ & 0.002 & 2007MNRAS.382.1133Z & 4286 & 313 & 0.073\\
   2 & DV Psc & 00:13:09.20 & +05:35:43.06 & Sec & 0.021 & 0.039 & 3614 &   8 & 0.002 & 2007MNRAS.382.1133Z & 3223 & 233 & 0.072\\
   3 & MU Cas & 00:15:51.56 & +60:25:53.64 & Pri & 0.021 & 0.012 & 14750 & 500 & 0.034 & 2004AJ....128.1840L & 13876 & 1321 & 0.095\\
   4 & MU Cas & 00:15:51.56 & +60:25:53.64 & Sec & 0.020 & 0.011 & 15100 & 500 & 0.033 & 2004AJ....128.1840L & 14569 & 1386 & 0.095\\
   5 & YZ Cas & 00:45:39.08 & +74:59:17.06 & Pri & 0.004 & 0.012 & 9200 & 300 & 0.033 & 1981ApJ...251..591L & 8937 & 558 & 0.062\\
   6 & YZ Cas & 00:45:39.08 & +74:59:17.06 & Sec & 0.007 & 0.015 & 6700 & 250 & 0.037 & 1981ApJ...251..591L & 6832 & 428 & 0.063\\
\hline
  \end{tabular}
 }
\end{table*}

\begin{table*}
 \setcounter{table}{6} 
 \setlength{\tabcolsep}{1pt}
 \renewcommand{\arraystretch}{.7}
 \tiny{
  \begin{tabular}{clcccccccclccc}
   \hline
   &    &     &     &     &      &      & \multicolumn{4}{c}{Published temperatures}    & \multicolumn{3}{c}{This study}\\
   \hline
   ID &       Star &   $\alpha$(J2000) & $\delta$(J2000) & Com & $\Delta M/M$ & $\Delta R/R$ &    $T_{eff}$ & $\Delta T_{eff}$ & $\Delta T/T$ & \multicolumn{1}{c}{Ref.}&    $T_{eff}$ & $\Delta T_{eff}$ & $\Delta T/T$ \\
   \hline
   76 & GG Ori & 05:43:10.22 & +00:41:14.90 & Pri & 0.007 & 0.013 & 9950 & 200 & 0.020 & 2000AJ....120.3226T & 10648 & 666 & 0.063 \\
   77 & GG Ori & 05:43:10.22 & +00:41:14.90 & Sec & 0.007 & 0.014 & 9950 & 200 & 0.020 & 2000AJ....120.3226T & 10687 & 669 & 0.063 \\
   78 & V1031 Ori & 05:47:26.89 & -10:31:58.65 & Sec & 0.007 & 0.021 & 8400 & 500 & 0.060 & 1990A$\&$A...228..365A & 8733 & 835 & 0.096 \\
   79 & beta Aur & 05:59:31.72 & +44:56:50.76 & Pri & 0.011 & 0.006 & 9350 & 200 & 0.021 & 1994A$\&$A...291..777N & 8828 & 549 & 0.062 \\
   80 & beta Aur & 05:59:31.72 & +44:56:50.76 & Sec & 0.012 & 0.007 & 9200 & 200 & 0.022 & 1994A$\&$A...291..777N & 8847 & 551 & 0.062 \\
   81 & V1388 Ori & 06:10:59.17 & +11:59:41.49 & Pri & 0.011 & 0.007 & 20500 & 500 & 0.024 & 2009AJ....137.3222W & 19495 & 1774 & 0.091 \\
\hline
  \end{tabular}
 }
\end{table*}

\begin{table*}
 \setcounter{table}{6} 
 \setlength{\tabcolsep}{1pt}
 \renewcommand{\arraystretch}{.7}
 \tiny{
  \begin{tabular}{clcccccccclccc}
   \hline
   &    &     &     &     &      &      & \multicolumn{4}{c}{Published temperatures}    & \multicolumn{3}{c}{This study}\\
   \hline
   ID &       Star &   $\alpha$(J2000) & $\delta$(J2000) & Com & $\Delta M/M$ & $\Delta R/R$ &    $T_{eff}$ & $\Delta T_{eff}$ & $\Delta T/T$ & \multicolumn{1}{c}{Ref.}&    $T_{eff}$ & $\Delta T_{eff}$ & $\Delta T/T$ \\
   \hline
   151 & 2MASS J09381349-0104281 & 09:38:13.51 & -01:04:27.90 & Sec & 0.028 & 0.017 & 4360 & 150 & 0.034 & 2011A$\&$A...527A..14H  & 4670 & 328 & 0.070 \\
   152 & DU Leo & 09:44:11.38 & +25:21:11.41 & Pri & 0.021 & 0.025 & $-$ & $-$ & $-$ & $-$ & 4823 & 341 & 0.071 \\
   153 & QX Car & 09:54:33.88 & -58:25:16.59 & Pri & 0.013 & 0.014 & 23800 & 500 & 0.021 & 1983A$\&$A...121..271A & 25874 & 2360 & 0.091 \\
   154 & QX Car & 09:54:33.88 & -58:25:16.59 & Sec & 0.014 & 0.015 & 22600 & 500 & 0.022 & 1983A$\&$A...121..271A & 25065 & 2287 & 0.091 \\
   155 & HS Hya & 10:24:36.77 & -19:05:32.96 & Pri & 0.006 & 0.005 & 6500 & 50 & 0.008 & 1997AJ....114.2764T & 6526 & 406 & 0.062 \\
   156 & HS Hya & 10:24:36.77 & -19:05:32.96 & Sec & 0.006 & 0.006 & 6400 & 50 & 0.008 & 1997AJ....114.2764T & 6473 & 403 & 0.062 \\
\hline
  \end{tabular}
 }
\end{table*}

\begin{table*}
 \setcounter{table}{6} 
 \setlength{\tabcolsep}{1pt}
 \renewcommand{\arraystretch}{.7}
 \tiny{
  \begin{tabular}{clcccccccclccc}
   \hline
   &    &     &     &     &      &      & \multicolumn{4}{c}{Published temperatures}    & \multicolumn{3}{c}{This study}\\
   \hline
   ID &       Star &   $\alpha$(J2000) & $\delta$(J2000) & Com & $\Delta M/M$ & $\Delta R/R$ &    $T_{eff}$ & $\Delta T_{eff}$ & $\Delta T/T$ &\multicolumn{1}{c}{Ref.}&    $T_{eff}$ & $\Delta T_{eff}$ & $\Delta T/T$ \\
   \hline
   226 & TX Her & 17:18:36.45 & +41:53:17.10 & Pri & 0.025 & 0.018 & 7534 & 200 & 0.027 & 2011NewA...16..498E & 7420 & 466 & 0.063 \\
   227 & TX Her & 17:18:36.45 & +41:53:17.10 & Sec & 0.021 & 0.021 & 6678 & 211 & 0.032 & 2011NewA...16..498E & 7170 & 452 & 0.063 \\
   228 & LV Her & 17:35:32.40 & +23:10:30.60 & Pri & 0.008 & 0.009 & 6060 & 150 & 0.025 & 2009AJ....138.1622T & 5988 & 373 & 0.062 \\
   229 & LV Her & 17:35:32.40 & +23:10:30.60 & Sec & 0.007 & 0.008 & 6030 & 150 & 0.025 & 2009AJ....138.1622T & 5963 & 371 & 0.062 \\
   230 & V624 Her & 17:44:17.25 & +14:24:36.24 & Pri & 0.006 & 0.010 & 8150 & 150 & 0.018 & 1984AJ.....89.1057P & 8078 & 504 & 0.062 \\
   231 & V624 Her & 17:44:17.25 & +14:24:36.24 & Sec & 0.007 & 0.014 & 7945 & 150 & 0.019 & 1984AJ.....89.1057P & 7671 & 480 & 0.063 \\
\hline     
  \end{tabular}
 }
\end{table*}

\begin{table*}
 \setcounter{table}{6} 
 \setlength{\tabcolsep}{1pt}
 \renewcommand{\arraystretch}{.7}
 \tiny{
  \begin{tabular}{clcccccccclccc}
   \hline
   &    &     &     &     &      &      & \multicolumn{4}{c}{Published temperatures}    & \multicolumn{3}{c}{This study}\\
   \hline
   ID &       Star &   $\alpha$(J2000) & $\delta$(J2000) & Com & $\Delta M/M$ & $\Delta R/R$ &    $T_{eff}$ & $\Delta T_{eff}$ & $\Delta T/T$ & \multicolumn{1}{c}{Ref.}&    $T_{eff}$ & $\Delta T_{eff}$ & $\Delta T/T$ \\
   \hline
   301 & V478 Cyg & 20:19:38.75 & +38:20:09.20 & Pri & 0.054 & 0.016 & 30550 & 1070 & 0.035 & 1991AJ....101..600P & 29246 & 2670 & 0.091 \\
   302 & V478 Cyg & 20:19:38.75 & +38:20:09.20 & Sec & 0.055 & 0.016 & 30550 & 1070 & 0.035 & 1991AJ....101..600P & 28824 & 2632 & 0.091 \\
   303 & MY Cyg & 20:20:03.39 & +33:56:35.02 & Pri & 0.017 & 0.009 & 7050 & 200 & 0.028 & 2009AJ....137.2949T & 7346 & 458 & 0.062 \\
   304 & MY Cyg & 20:20:03.39 & +33:56:35.02 & Sec & 0.017 & 0.009 & 7000 & 200 & 0.029 & 2009AJ....137.2949T & 7200 & 449 & 0.062 \\
   305 & V399 Vul & 20:25:10.80 & +21:29:18.84 & Pri & 0.011 & 0.005 & 19000 & 320 & 0.017 & 2012NewA...17..215C & 18318 & 1667 & 0.091 \\
   306 & V399 Vul & 20:25:10.80 & +21:29:18.84 & Sec & 0.006 & 0.025 & 18250 & 520 & 0.028 & 2012NewA...17..215C & 17697 & 1695 & 0.096 \\
\hline
  \end{tabular}
 }
\end{table*}

\pagebreak

\end{document}